\documentclass[journal]{IEEEtran}

\usepackage{graphicx}
\usepackage{subfigure}
\usepackage{amstext}
\usepackage{bm}
\usepackage{cite}
\usepackage{amsmath}
\usepackage{epstopdf}
\usepackage{xfrac}
\usepackage{amsfonts,amssymb}
\usepackage{color}
\definecolor{kkb}{RGB}{0,0,0}
\usepackage{subfigure}
\usepackage{algorithm}
\usepackage{algorithmic}
\usepackage{algorithm}
\usepackage{algorithmic}
\renewcommand{\algorithmicrequire}{\textbf{Initialization:}}   %modify the title
\newcommand{\INIT}{\item[\algorithmicrequire]}

\newcommand{\algorithmicinput}{\textbf{Input:}}
\newcommand{\INPUT}{\item[\algorithmicinput]}

\newcommand{\algorithmicoutput}{\textbf{Output:}}
\newcommand{\OUTPUT}{\item[\algorithmicoutput]}

\newtheorem{remark}{Remark}

\begin{document}

\title{Codebook-Based Beam Tracking for Conformal Array-Enabled UAV MmWave Networks}

%\markboth{Journal of \LaTeX\ Class Files,~Vol.~14, No.~8, August~2015}

\author{Jinglin Zhang,~\IEEEmembership{Student Member,~IEEE,}
Wenjun Xu,~\IEEEmembership{Senior Member,~IEEE,}
Hui Gao,~\IEEEmembership{Senior Member,~IEEE,}\\
Miao Pan,~\IEEEmembership{Senior Member,~IEEE,}
Zhu Han,~\IEEEmembership{Fellow,~IEEE,}
and~Ping Zhang,~\IEEEmembership{Fellow,~IEEE}
\thanks{Copyright (c) 20xx IEEE. Personal use of this material is permitted. However, permission to use this material for any other purposes must be obtained from the IEEE by sending a request to pubs-permissions@ieee.org.}
\thanks{This work was supported in part by the National Key R\&D Program of China {(No. 2019YFC1511302)}, the National Natural Science Foundation of China under Grant {(No. 61871057)}, and the Fundamental Research Funds for the Central Universities(2019XD-A13), in part by the U.S. National Science Foundation under grants US CNS-1350230 (CAREER), CNS-1702850, CNS-1801925, and CNS-2029569, in part by US Multidisciplinary University Research Initiative 18RT0073, NSF EARS-1839818, CNS1717454, CNS-1731424, and CNS-1702850, and in part by BUPT Excellent Ph.D. Students Foundation (CX2020202).}
\thanks{J. Zhang and W. Xu are with key Lab of Universal Wireless Communications, Ministry of Education,
Beijing University of Posts and Telecommunications, Beijing,100876 China.~{(E-mail: {zjinglin, wjxu}@bupt.edu.cn)}~{(Corresponding author: Wenjun Xu)}.}
\thanks{H. Gao is with Key Laboratory of Trustworthy Distributed Computing and Service, Ministry of Education,
Beijing University of Posts and Telecommunications, Beijing,100876 China.~(E-mail: huigao@bupt.edu.cn)}
\thanks{M. Pan is with the Department of Electrical and Computer Engineering,
University of Houston, Houston, TX 77204 USA.}
\thanks{Z. Han is with the University of Houston, Houston, TX 77004 USA, and also with the Department of Computer Science and Engineering, Kyung Hee University, Seoul, South Korea, 446-701. }
\thanks{P. Zhang is with State Key Laboratory of Networking and Switching Technology,
Beijing University of Posts and Telecommunications, Beijing, 100876 China.}
}

\maketitle
\vspace{-0.5cm}
\begin{abstract}
Millimeter wave (mmWave) communications can potentially meet the high data-rate requirements of unmanned aerial vehicle (UAV) networks. However, as the prerequisite of mmWave communications, the narrow directional beam tracking is very challenging because of the three-dimensional (3D) mobility and attitude variation of UAVs. Aiming to address the beam tracking difficulties, we propose to integrate the conformal array (CA) with the surface of each UAV, which enables the full spatial coverage and the agile beam tracking in highly dynamic UAV mmWave networks. More specifically, the key contributions of our work are three-fold. 1) A new mmWave beam tracking framework is established for the CA-enabled UAV mmWave network. 2) A specialized hierarchical codebook is constructed to drive the directional radiating element (DRE)-covered cylindrical conformal array (CCA), which contains both the angular beam pattern and the subarray pattern to fully utilize the potential of the CA. 3) A codebook-based multiuser beam tracking scheme is proposed, where the Gaussian process machine learning enabled UAV position/attitude predication is developed to improve the beam tracking efficiency in conjunction with the tracking-error aware adaptive beamwidth control. Simulation results validate the effectiveness of the proposed codebook-based beam tracking scheme in the CA-enabled UAV mmWave network, and demonstrate the advantages of CA over the conventional planner array in terms of spectrum efficiency and outage probability in the highly dynamic scenarios.
\end{abstract}


% \fi
\IEEEpeerreviewmaketitle

\section{Introduction}
Unmanned aerial vehicles (UAVs) have many critical applications in civilian and military areas~\cite{zeng2019cellular,zeng2019accessing}, such as fire fighting, plant protection, remote monitoring, etc. Therein, a UAV acts as an Internet of Things (IoT) node or a peripheral node to assist IoT. For instance, it is very attractive to deploy UAVs to assist IoT with satisfying the requirements of high-rate and low-latency data transmission~\cite{55,30}. \textcolor{black}{Moreover, UAVs equipped with IoT sensors can be regarded as part of IoT to perform various missions flexibly~\cite{35}.} In many mission-driven scenarios, e.g., gathering and relaying IoT data, cooperative radio transmission, environment sensing, traffic flow monitoring, etc., multiple UAVs are often teamed to collaboratively accomplish the designated missions, in which real-time sensing/monitoring information or high-definition video transmission are usually necessitated~\cite{2}. \textcolor{black}{Therefore, the high-rate data transmission among UAVs is of great importance to the development of IoT with large-scale video data or high-definition image data transmission requirements. In addition, UAVs can also be used as flying base-stations or mobile relay backhaul nodes to provide on-the-fly high-capacity communication links for the emergency coverage of IoT devices~\cite{28,32}.}

In such mission-driven UAV networks, high-data-rate inter-UAV communications play a pivotal role. MmWave band has abundant spectrum resource, and is considered as a potential avenue to support high-throughput data transmission for UAV networks~\cite{1,zhang2019research,28}. \textcolor{black}{If the Line-of-Sight (LoS) propagation is available, mmWave communication can achieve kilometer-level communication range and gigabits-persecond data rate~\cite{58}, which can support UAV networks in many scenarios~[6],~[7],~[9].} However, there are critical challenges to achieve reliable mmWave communications for UAV networks. Specifically, a UAV maintains three-dimensional or full-spatial mobility with very high dynamic, and thus the angle-of-arrival (AOA) of communication signal always varies over time in all directions. To this end, a powerful antenna array is important to offer full-spatial coverage capability and facilitate the mmWave link maintenance for UAV networks. The uniform linear array (ULA) and uniform planar array (UPA) are widely adopted in the existing studies on mmWave communication and networking~\cite{13,14,15,16}. However, their coverage capabilities are often confined within a two-dimensional space and a half three-dimensional space. Therefore, conventional ULA and UPA can only attain a limited coverage within a fraction of the full space, causing high communication outage probability in highly dynamic UAV mmWave networks.

\textcolor{black}{When considering UAV communications with UPA or ULA, a UAV is typically modeled as a point in space without considering its size and shape. Actually, the size and shape can be utilized to support more powerful and effective antenna array.} Inspired by this basic consideration, the conformal array (CA)~\cite{4} is introduced to UAV communications. A CA is usually in a shape of cylindrical or spherical conforming to a predefined surface, e.g., a part of an airplane or UAV, and can reap full spatial coverage with proper array designs. \textcolor{black}{Compared with surface-mounted multiple UPAs, a CA, conforming to the surface of a UAV, can compact the UAV design, reduce the extra drag and fuel consumption, and also facilitate an array of a larger size~\cite{4}. Furthermore, directional radiating elements (DREs) are commonly integrated with antenna array to enhance the beamforming ability~\cite{4,8,9}. In such a case, the coverage capability of CA is far stronger than that of UPA and ULA via proper array designs, due to the exploitation of size and shape. Specifically, a CA can enable the potential to enlarge (roll up) the surface of antenna array.} This advantage not only achieves a larger array gain to combat path-loss but also sustains full-spatial transmitting/receiving to facilitate fast beam tracking for mobile UAV mmWave networks~\cite{6}. Note that in mission-driven UAV networks, agile and robust beam tracking is very challenging yet critical for inter-UAV mmWave communications~\cite{zhang2019research}, because UAV position and attitude may vary very fast. By carefully exploiting the CA's full spatial transmission/reception property, the stringent constraints on beam tracking for highly dynamic moving UAVs can be relieved considerably. So far, however, the CA-enabled UAV mmWave network is almost untouched in the literature. Regarding the mmWave CA, there are only a few recent works on the radiation patterns and beam scanning characteristics~\cite{7} and the performance evaluation of CA-based beamforming for \emph{static mmWave cellular networks}~\cite{5}. These works validate the potential advantage of CA in the static mmWave networks, which are not applicable to \emph{mobile UAV mmWave networks}.

For both static and mobile mmWave networks, codebook design is of vital importance to empower the feasible beam tracking and drive the mmWave antenna array for reliable communications~\cite{26,27}. Recently, ULA/UPA-oriented codebook designs have been proposed for mmWave networks, which include the codebook-based beam tracking and channel estimation methods. For example, considering the ULA with omnidirectional radiating elements (REs), the hierarchical-codebook-based subarray and antenna deactivating strategies are proposed to achieve efficient beam training for single-user scenarios~\cite{13,11}. The multiuser downlink beam training algorithms regarding the ULA are proposed with the multi-resolution codebook designs for partially-connected~\cite{12} and fully-connected~\cite{16} hybrid structures, respectively. However, extending the aforementioned works to the CA is not straightforward. The reasons are as follows: When the commonly-adopted DRE is integrated with CA, the limited radiation range of DREs is no longer the same and each is affected by the DRE's location on CA, as the DRE-covered array plane is rolled up. The determined radiation direction of CA is only within a part of DREs' radiation range. This observation indicates that only a part of the DREs or some specific subarrays need to be activated with reference to the AOA or angle of departure (AOD) of transceivers.
Therefore, the dynamic subarray localization and activation are very coupled and critical for the efficient utilization of the DRE-covered CA. Note that conventional ULA/UPA-oriented codebook designs mainly focus on the beam direction/width controlling via the random-like subarray activation/deactivation without specific subarray localization. In contrast, the codebook design for DRE-covered CA should emphasize the location of the activated subarray to achieve the promise of full-spatial coverage of the CA in UAV networks. Nevertheless, such work is still missing now in the literature. These points mentioned above motivate us to study a new beam tracking framework with the well-tailored codebook for CA-enabled UAV mmWave networks.

In this paper, we consider a dynamic mission-driven UAV network with UAV-to-UAV mmWave communications, wherein multiple transmitting UAVs (t-UAVs) simultaneously transmit to a receiving UAV (r-UAV). \textcolor{black}{In such a scenario, we focus on inter-UAV communications in UAV networks, and the UAV-to-ground communications are not involved.} In particular, each UAV is equipped with a cylindrical conformal array (CCA), and a novel-codebook-based mmWave beam tracking scheme is proposed for such a highly dynamic UAV network. More specifically, the codebook consists of the codewords corresponding to various subarray patterns and beam patterns. Based on the joint UAV position-attitude prediction, an efficient codeword selection scheme is further developed with tracking error (TE) awareness, which achieves fast subarray activation/partition and array weighting vector selection. It is verified that our proposed scheme achieves a higher spectrum efficiency, lower outage probability and stronger robustness for inter-UAV mmWave communications. In summary, the key contributions of this paper are listed as follows.
 	\begin{itemize}
    \item \emph{The first study on the beam tracking framework for CA-enabled UAV mmWave networks}. We propose an overall beam tracking framework to exemplify the idea of the DRE-covered CCA integrated with UAVs, and reveal that CA can offer full-spatial coverage and facilitate beam tracking, thus enabling high-throughput inter-UAV data transmission for mission-driven UAV networking. To the best of our knowledge, this is the first work on the beam tracking framework for CA-enabled UAV mmWave networks.
    \item \emph{The specialized codebook design of the DRE-covered CCA for multi-UAV mobile mmWave communications}. Under the guidance of the proposed framework, a novel hierarchical codebook is designed to encompass both the subarray patterns and beam patterns. The newly proposed CA codebook can fully exploit the potentials of the DRE-covered CCA to offer full spatial coverage. Moreover, the corresponding codeword selection scheme is also carefully designed to facilitate fast multi-UAV beam tracking/communication in the considered CA-enabled UAV mmWave network.
    \item \emph{The CCA codebook-based multi-UAV beam tracking scheme with TE awareness}. Based on the designed codebook, by exploiting the Gaussian process (GP) tool, both the position and attitude of UAVs can be fast tracked for fast multiuser beam tracking along with dynamic TE estimation. Moreover, the estimated TE is leveraged to direct the selection of a proper codeword, which inspires an optimized subarray activation and beamwidth control towards the better performance with TE awareness.
  \end{itemize}

Note that there exist some mobile mmWave beam tracking schemes exploiting the position or motion state information (MSI) based on conventional ULA/UPA recently. For example, the beam tracking is achieved by directly predicting the AOD/AOA through the improved Kalman filtering~\cite{23}, however, the work of~\cite{23} only targets at low-mobility scenarios. For vehicle networks, the position-assisted beam tracking methods are proposed by~\cite{20} and~\cite{21}. Nevertheless, the impact of the attitude changes of vehicles on the beam tracking is not involved. The research work on the beam tracking for UAVs with mmWave communications is still rare. The authors in~\cite{19} and~\cite{22} consider the UAV-to-ground and UAV-to-satellite mmWave communications, respectively, with fast beam tracking by estimating the position and attitude of UAVs. However, previous schemes cannot be readily extended to UAV-to-UAV mmWave communications, where both transmitter and receiver are fast moving with quick attitude variations. Recently, we propose a position-attitude-prediction-based beam tracking scheme for the UAV-to-UAV mmWave communication with conventional UPA~\cite{25}, which is an initial attempt to address the beam tracking challenge in UAV mmWave networks by adopting the GP tool. In a nutshell, all the aforementioned work~\cite{23,20,21,19,22,25} is based on conventional ULA/UPA, and there is no existing work on the beam tracking solution for CA-enabled UAV mmWave networks. Moreover, the aforementioned work~\cite{23,20,21,19,22,25} does not consider the TE-aware robust design, which might be much beneficial for highly dynamic UAV networks.

\newcounter{mytempeqncnt}
\begin{figure*}[!t]
\normalsize
\setcounter{mytempeqncnt}{\value{equation}}
\setcounter{equation}{0}
\begin{eqnarray}
\label{steering}
\boldsymbol{A}({{\alpha }},{{\beta }})\!&=&\!\left[{{e}^{j\frac{2\pi }{\lambda_c }(\frac{(1-M)d_\text{cyl}}{2}\cos {{\beta }}+R_\text{cyl}\sin\frac{(1-N)\Delta\phi_\text{c}}{2}\sin {{\alpha }}\sin {{\beta }}+R_\text{cyl}\cos\frac{(1-N)\Delta\phi_\text{c}}{2}\cos\alpha_t\sin\beta)}},\ldots , \right.\\
&&{{e}^{j\frac{2\pi }{\lambda_c }(\frac{(M+1-2m)d_\text{cyl}}{2}\cos {{\beta }}+R_\text{cyl}\sin\phi_c(n)\sin {{\alpha }}\sin {{\beta }}+R_\text{cyl}\cos\phi_{c}(n)\cos\alpha\sin\beta)}},\nonumber \\
&&\left.{\ldots},{{e}^{j\frac{2\pi }{\lambda_c }(\frac{(M-1)d_\text{cyl}}{2}\cos {{\beta }}+R_\text{cyl}\sin\frac{(N-1)\Delta\phi_\text{c}}{2}\sin {{\alpha }}\sin {{\beta }}+R_\text{cyl}\cos\frac{(N-1)\Delta\phi_\text{c}}{2}\cos\alpha\sin\beta)}}\right]^{T},\nonumber
\end{eqnarray}
\setcounter{equation}{\value{mytempeqncnt}}
\hrulefill
\vspace*{4pt}
\end{figure*}

The rest of this paper is as follows. In Section~\ref{Sec2}, the system model is introduced. In Section~\ref{Sec3}, the CCA codebook design and the codebook-based joint subarray partition and AWV selection algorithms are proposed. Next, the TE-aware codebook-based beam tracking with 3D beamwidth control is further proposed in Section~\ref{Sec4}. Simulation results are given in Section~\ref{Sec5}, and finally Section~\ref{Sec6} concludes this paper.
%notation

%\IEEEPARstart{T}{his} demo file is intended to serve as a ``starter file''
%for IEEE journal papers produced under \LaTeX\ using
%IEEEtran.cls version 1.8b and later.
%I wish you the best of success.



%\subsection{Subsection Heading Here}
%Subsection text here.

%\IEEEpubidadjcol

%\subsubsection{Subsubsection Heading Here}
%Subsubsection text here.
\section{System Model}
\label{Sec2}
A CCA-enabled UAV mmWave network is considered in this paper. Here, we first establish the DRE-covered CCA model in Section~\ref{Sec2.1}. Then the system setup of the considered UAV mmWave network is described in Section~\ref{Sec2.2}. Finally, the beam tracking problem for the CA-enabled UAV mmWave network is modeled in Section~\ref{Sec2.3}.

\subsection{Conformal Array Model}
\label{Sec2.1}
\begin{figure}[!t]
\centering
\includegraphics[height=1.4in,width=2.9in]{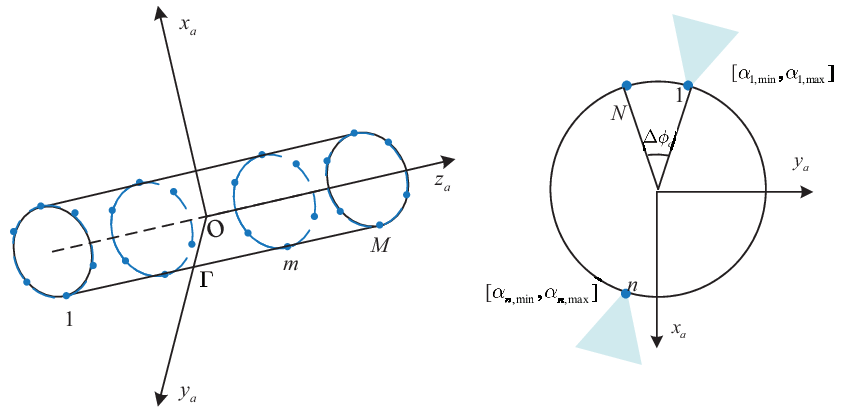}
\caption{The DRE-covered CCA in 3D view and top-view. }
\label{figsim1}
\end{figure}

 The CCA is the conformal array with a cylindrical shape that can conform some parts of UAV's configuration such as the fuselage. A CCA contains $M\times N$ elements, ${{M}}\gg 1,{{N}}\gg 1$, which are placed along the cylindrical shape $\Gamma$ with $N$ and $M$ elements on the $xy$-plane and the $z$-axis, respectively, as shown in Fig.~\ref{figsim1}. The elements of CCA on the $xy$-plane form a circular array and the elements of CCA on the $z$-axis form a uniform line array.
%Moreover, each DRE is equipped with a phase shifter. The fully-connected hybrid architecture is adopted in the considered system. In a certain slot, the partly-connected architecture forms with the subarray partitioning dynamically.
According to \cite{5,xu2008pattern}, the steering vector is given by~(\ref{steering}), where $\alpha$, $\beta$ and $R_\text{cyl}$ are the azimuth angle, the elevation angle and the radius of the cylinder, respectively. The inter-element distance on the
$z$-axis $d_\text{cyl}=\frac{\lambda_c}{2}$, where $\lambda_c$ is the carrier wavelength. The inter-element distance on the $xy$-plane is also set as $d_{xy}=\frac{\lambda_c}{2}$. $\phi_n=\phi_\text{c}(n)=\frac{(2n-1-N)\Delta\phi_\text{c}}{2}$ is the angular position of the $n$-th element on the $xy$-plane and $\Delta\phi_\text{c}=\frac{2\pi}{N}$ is the corresponding inter-element distance in the angular domain.
It is assumed that the DREs are modeled as ideal directional elements with the angular domain radiation coverage $\Delta\alpha$ and $\Delta\beta$~\cite{SpatiallySparse}. More specifically, as shown by Fig.~\ref{figsim1}, each DRE has the azimuth angle coverage, $[{{\alpha }_{n,\min}},{{\alpha }_{n,\max}}]$ with $\Delta\alpha={{\alpha }_{n,\max}}-{{\alpha }_{n,\min}}$. The elevation angle covered by a DRE is denoted by $[{{\beta }_{m,\min}},{{\beta }_{m,\max}}]$ with $\Delta\beta={{\beta }_{m,\max}}-{{\beta }_{m,\min}}$. In the CCA coordinate frame shown in Fig.~\ref{figsim1}, the azimuth angle coverage of the $n$-th element on the $xy$-plane is given by
\begin{eqnarray}
 \setcounter{equation}{2}
\label{DREcover}
\begin{aligned}
&{{\alpha }_{n,\min}}=\phi_n-\frac{\Delta\alpha}{2}+2l\pi, l\in\mathbb{Z},\\
&{{\alpha }_{n,\max}}=\phi_n+\frac{\Delta\alpha}{2}+2l\pi, l\in\mathbb{Z},
\end{aligned}
\end{eqnarray}
 and the elevation angle coverage of the $m$-th element on $z$-axis is given by
 \begin{eqnarray}
\label{DREcoverbeta}
\begin{aligned}
&{{\beta }_{m,\min}}=-\frac{\Delta\beta}{2}+2l\pi, l\in\mathbb{Z},\\
&{{\beta }_{m,\max}}=\frac{\Delta\beta}{2}+2l\pi, l\in\mathbb{Z}.
\end{aligned}
\end{eqnarray}
\textcolor{black}{For the CCA-enabled UAV mmWave network, the array size is usually large and the corresponding inter-element distance $\Delta\phi$ is small. Therefore, it is assumed that $\Delta\alpha$ and $\Delta\beta$ satisfy $\Delta\phi_{\text{c}}\leq\Delta\alpha$ and $\Delta\beta=\pi$ to ensure that the DRE-covered CCA covers the full angular domain. }
%Meanwhile, the radius of the cylinder $R_\text{cyl}$ and $N$ satisfy $R_\text{cyl}}=*N*\lambda$

\begin{figure}[!t]
\centering
\includegraphics[height=1.3in,width=2.3in]{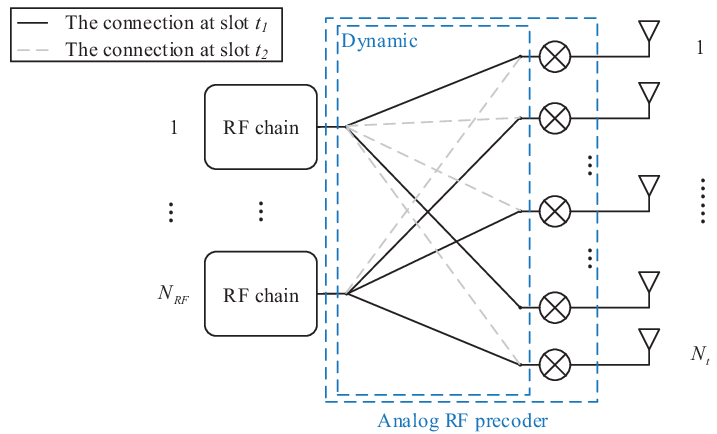}
\caption{\textcolor{black}{The analog RF precoder structure with dynamic subarrays.} }
\label{precoder}
\end{figure}

\textcolor{black}{The analog precoding architecture adopted for DRE-covered CCA is shown in Fig.~\ref{precoder}~\cite{14}, which tunes the partially-connected precoding architecture by adapting the connection between the RF chains and the antenna elements to the channel variation and forming dynamic subarrays. For a fixed time slot, the precoding architecture is the conventional partially-connected precoding. For different time slots, the connection changes mainly depend on the variations of the AOA/AOD caused by the movement of UAVs. }
%\begin{eqnarray}
%&&\boldsymbol{A}({{\alpha }_t},{{\beta }_{t}})=\\
%&&\left[{{e}^{j\frac{2\pi d}{\lambda_c }(R_{cyl}\cos\frac{(1-N)}{2}\cos {{\beta }_{t}}+R_{cyl}\sin(1-N)\sin {{\alpha }_{t}}\sin {{\beta %}_{t}}+\frac{(1-M)d_{cyl}}{2}\cos(\alpha_t)\sin(\beta_t))}},\ldots , \right.\\
%&&{{e}^{j\frac{2\pi d}{\lambda_c }(R_{cyl}\cos\frac{(1-N)}{2}\cos {{\beta }_{t}}+R_{cyl}\sin(1-N)\sin {{\alpha }_{t}}\sin {{\beta %}_{t}}+\frac{(1-M)d_{cyl}}{2}\cos(\alpha_t)\sin(\beta_t))}}\\
%&&\left.{\ldots},{{e}^{j\frac{2\pi d((M-1)\sin {{\alpha }_{t}}\sin {{\beta }_{t}}+(M-1)\cos {{\alpha }_{t}}\sin {{\beta }_{t}})}{\lambda_c }}}\right]^{T},\nonumber
%\end{eqnarray}

\subsection{System Setup}
\label{Sec2.2}
\begin{figure}[!t]
\centering
\includegraphics[scale=0.8,width=2in]{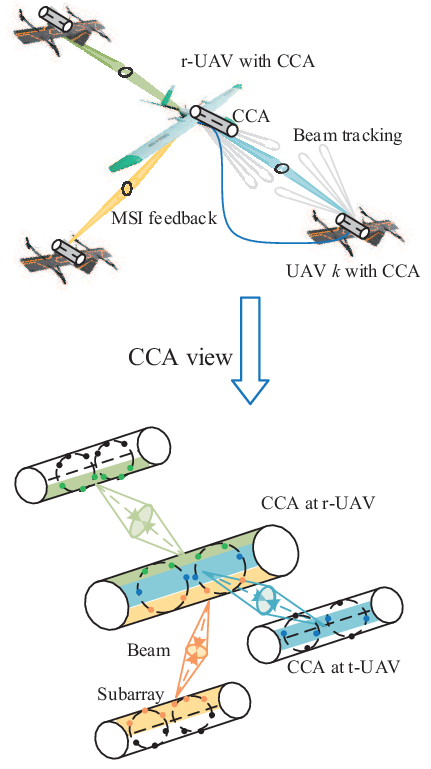}
\caption{The considered CC-enabled UAV mmWave network consists of a r-UAV and multiple t-UAVs. UAV position-attitude prediction is performed to obtain the future motion state information (MSI) before next information feedback. The CCA and the beam are shown in detail in the CCA view.}
\label{figsim2}
\vspace{-15pt}
\end{figure}

\begin{table}[!t]
\renewcommand{\arraystretch}{1.3}
\caption{SUMMARY OF MAIN NOTATIONS}
\label{table1}
\centering
\begin{tabular}{c||c}
\hline
\bfseries Symbol & \bfseries Definition\\
\hline\hline
$M$ & The number of CCA elements on the $z$-axis \\
\hline
$N$ & The number of CCA elements on the $xy$-plane \\
\hline
$M_{\text{act}}$ & The number of activated subarray elements on the $z$-axis \\
\hline
$N_{\text{act}}$ & The number of activated subarray elements on the $xy$-plane \\
\hline
$N_{\text{RF}}$ & The number of RF chains \\
\hline
$(m,n)$ & The position of the DRE on CCA \\
\hline
$(m_s,n_s)$ & The layer index in the CCA codebook \\
\hline
$(m_c,n_c)$ & The position of center element of the subarray \\
\hline
$\Delta\phi_c$ & The angular distance between the elements on $xy$-plane\\
\hline
$\boldsymbol{\Lambda}$ & The antenna element gain\\
\hline
$\boldsymbol{v}$ & The codeword (AWV) in 3D CCA codebook\\
\hline
$\mathcal{S}$ & The activated subarray\\
\hline
$\boldsymbol{X}_{t(r)}$ & The position of t(r)-UAV\\
\hline
$\boldsymbol{\Theta}_{t(r)}$ & The attitude of t(r)-UAV\\
\hline
\end{tabular}
\end{table}
A mission-driven UAV network consisting of a leading UAV (LUAV) and several following UAVs (FUAVs) is considered. The FUAVs follow the LUAV to perform tasks and transmit data to it, and the LUAV has strong ability to communicate with and forward the aggregated data to the remote ground station. Since real-time and high data rate information transmission is usually required by these tasks, mmWave communication is adopted in the UAV network to meet the demands. As shown in Fig.~\ref{figsim2}, the considered UAV mmWave network is composed of $K$ t-UAVs (FUAVs) and one r-UAV (LUAV), each t-UAV and the r-UAV are equipped with a CCA with ${{M_\text{t}}}\times {{N_\text{t}}}$ and ${{M_\text{r}}}\times {{N_\text{r}}}$ DREs, respectively. It is assumed that only analog beamforming is considered and only one radio frequency (RF) chain is needed for t-UAVs. The r-UAV is equipped with $N_\text{RF}$ RF chains with $N_\text{RF}>K$. Denote
$\boldsymbol{H}_k\in\mathbb{C}^{M_{\text{t}}N_{\text{t}}\times {M_{\text{r}}N_{\text{r}}}}, k=1,\ldots,K,$ as the multiple-input multiple-output (MIMO) channel between the $k$-th t-UAV and r-UAV. For an $M_\text{t}N_\text{t}\times 1$ unit-norm analog transmit beamforming vector, i.e., the transmit antenna weight vector (AWV) $\boldsymbol{f}_k$, and an $M_\text{r}N_\text{r}\times 1$ unit-norm analog combing vector, i.e., the receive AWV, $\boldsymbol{w}_k$, the received signal of the $k$-th t-UAV at r-UAV is expressed as
\begin{eqnarray}
r_k=\sqrt{p_k}\boldsymbol{w}_{{k}}^{H}{{\boldsymbol{H}}_{{k}}}{{\boldsymbol{f}}_{{k}}}s_k+\boldsymbol{w}_{{k}}^{H}\sum_{i\neq k}{\sqrt{p_i}\boldsymbol{H}_i\boldsymbol{f}_is_i}+\boldsymbol{w}_{{k}}^{H}{{\boldsymbol{n}}_{{}}},
\label{signal}
\end{eqnarray}
where $s_k$ is the transmitted signal of the $k$-th t-UAV with $\mathbb{E}\{|s_k|^2\}=1$, $p_k$ is the transmit power of the $k$-th t-UAV, and $\boldsymbol{n}$ is the $M_\text{r}N_\text{r}\times 1$ noise vector, whose elements are independent and identically distributed, and obey $\mathcal{CN}(0,\sigma_n^2)$, where $\sigma_n^2$ is the noise variance. The rich scattering environment rarely appears in the UAV mmWave communication since there are few scatterers in the air. \textcolor{black}{At the appropriate altitude, the airspace is relatively open, and the blockage hardly happens. Hence, the LoS propagation is dominated in the inter-UAV communication for our considered UAV mmWave networks~\cite{2}.}
%At the appropriate altitude, the mmWave UAV-to-UAV channel is assumed to consist of only light-of-sight (LOS) propagation
%It is not considered here that the LOS channel is blocked by other UAVs.
Therefore, the channel matrix $\boldsymbol{H}$ from the t-UAV to r-UAV with DRE-covered CCA is given by~\cite{22, SpatiallySparse}
\begin{eqnarray}
\label{channel}
\begin{aligned}
\boldsymbol{H}_k(t)\!&=\!\frac{h_0}{D_k^{-\gamma}(t)}(\boldsymbol{\Lambda}_{k}(\alpha^{\text{r}}_{k}(t),\beta^{\text{r}}_{k}(t))\circ\boldsymbol{A}(\alpha^{\text{r}}_{k}(t),\beta_{k}^{\text{r}}(t)))\\
&(\boldsymbol{\Lambda}_{k}(\alpha_{k}^{\text{t}}(t),\beta_{k}^{\text{t}}(t))\circ\boldsymbol{A}(\alpha_{k}^{\text{t}}(t),\beta_{k}^{\text{t}}(t)))^H,
\end{aligned}
\end{eqnarray}
where $\gamma $ denotes the path-loss exponent, $D_k$ denotes the distance between the $k$-th t-UAV and r-UAV, $h_0$ is the complex channel gain and $\circ$ represents Hadamard product.
$\boldsymbol{A}(\alpha_{k}^{\text{r}},\beta_{k}^{\text{r}})$ and $\boldsymbol{A}(\alpha_{k}^{\text{t}},\beta_{k}^{\text{t}})$ are the normalized transmitting and receiving array response vectors for the $k$-th t-UAV at the azimuth (elevation) angles of arrival (AOAs) and departure (AODs) $\alpha_{k}^{\text{r}}(\beta_{k}^{\text{r}})$, ${{\alpha }_{k}^{\text{t}}}({{\beta }_{k}^{\text{t}}})$, respectively. $\boldsymbol{\Lambda}_{k}(\alpha_{k}^{\text{r}},\beta_{k}^{\text{r}})$ and $\boldsymbol{\Lambda}_{k}(\alpha_{k}^{\text{t}},\beta_{k}^{\text{t}})$ are the antenna element gains of the CCA at the azimuth (elevation) AOAs and AODs, respectively. The DRE is considered as the ideal sectored element~\cite{SpatiallySparse} and hence the element gain is given by
\begin{eqnarray}
{{\left[ {{\boldsymbol{\Lambda}}_{k}}\left( {{\alpha }^{\text{t}}_{k}},{{\beta }_{k}^{\text{t}}} \right) \right]}_{(m,n)}}=\left\{
\begin{IEEEeqnarraybox}[\relax][c]{ll}
 &1,\ \ \ \forall {{\alpha }_{k}^{\text{t}}}\in [{{\alpha }_{m,\min }},{{\alpha }_{m,\max }}],\\
 &\ \ \ \ \ \forall {{\beta }_{k}^{\text{t}}}\in [{{\beta}_{n,\min }},{{\beta }_{n,\max }}], \\
 &0,\ \ \ \ \ \ \ \ \ \ \ \ \ \ \ \ \ \text{otherwise.}
\end{IEEEeqnarraybox}\right.
\end{eqnarray}
\subsection{Beam Tracking Problem Formulation with CCA}
\label{Sec2.3}
For the LOS channel, the AOAs and AODs in (\ref{channel}) are mainly determined by the position and attitude of the t-UAVs and r-UAV.
%Ҫ��Ҫ��
Given the received signal in (\ref{signal}), the signal-to-interference-plus-noise ratio (SINR) \footnote{\textcolor{black}{%When the t-UAVs are close to each other, their transmit signals are likely to have similar AOAs at the r-UAV and the combiners cannot separate them correctly. Therefore, the inter-UAV interference exists in the considered mmWave UAV network.
When the transmit signals of t-UAVs have similar AOAs at the r-UAV, the combiners may not separate them correctly and the inter-UAV interference exists in the UAV networks. Due to the constraint on the minimum distance between UAVs, the inter-UAV interference is not the main factor affecting the sum SE and can be neglected, which will be shown in our simulations.}}
of the $k$-th t-UAV at slot $t$ is given by
\begin{eqnarray}
\label{SINR}
\gamma_k(t)\!=\!\frac{\left|\sqrt{p_{k}(t)}\boldsymbol{w}_{k}(t)^H\boldsymbol{H}_{k}(t)\boldsymbol{f}_{k}(t)\right|^2}{\sum\limits_{i\neq k}\left|\sqrt{p_{i}(t)}\boldsymbol{w}_{k}(t)^H\boldsymbol{H}_{i}(t)\boldsymbol{f}_{i}(t)\right|^2\!+\!\boldsymbol{n}_{\sigma}(t)},
\end{eqnarray}
where $\boldsymbol{n}_{\sigma}(t)=\sigma^2\boldsymbol{w}_{k}(t)^H\boldsymbol{w}_{k}(t)$. The sum spectral efficiency (SE) of the mmWave UAV network at slot $t$ is given by
\begin{eqnarray}
\label{SE}
R(t)=\sum_{k\in \mathcal{K}}\log(1+\gamma_k(t)),
\end{eqnarray}
Aiming to maximize the SE at slot $t$, the optimal beamforming and combining vector $\boldsymbol{f}_k(t)$ and $\boldsymbol{w}_k(t)$, which are mainly determined by the AOAs $\alpha_{k}^{\text{r}}(t)(\beta_{k}^{\text{r}}(t))$ and AODs ${{\alpha }_{k}^{\text{t}}}(t)({{\beta }_{k}^{\text{t}}}(t))$ at slot $t$, shall be carefully designed by efficiently solving the following problem
\begin{eqnarray}
\begin{aligned}\label{P0} & \underset{\boldsymbol{f}_{{k}},\thinspace\boldsymbol{w}_{{k}}}{\text{max}} &  & R(t)\\
 & \text{subject to} &  & \left\Vert \boldsymbol{f}_{{k}}\right\Vert =1,\\
 &  &  & \left\Vert \boldsymbol{w}_{{k}}\right\Vert =1.
\end{aligned}
\end{eqnarray}
Note that directly solving the above beam tracking problem is very challenging, especially in the considered highly dynamic UAV mmWave network. Therefore, developing new and efficient beam tracking solution for the CA-enabled UAV mmWave network is the major focus of our work. Recall that several efficient codebook-based beam training and tracking schemes have been proposed for conventional mmWave network with uniform ULA and UPA \cite{26,27}. These prior works inspire us to propose a specialized new codebook design and the corresponding codeword selection/processing strategy that can drive the CCA to achieve fast beam tracking in the highly dynamic UAV mmWave network. To this end, the properties of the CCA should be exploited in the design of the codebook, which are briefly discussed as follows.

\emph{\textbf{Activated Subarray with Limited DREs}}: As shown in Fig.~\ref{figsim1}, given a certain azimuth angle, there are limited DREs that can be activated. Due to the directivity, the DREs of the CCA subarray at different positions are anisotropic, and this phenomenon is different from the UPA. If an inappropriate subarray is activated, the beam angle may go beyond the radiation range of certain subarray elements, degrading the beam gain and SE.

\emph{\textbf{Multiuser-resultant Receiver Subarray Partition}}: As shown in Fig.~\ref{figsim2}, the r-UAV needs to activate multiple subarrays to serve multiple t-UAVs at the same time. Assuming that an element can not be contained in different subarrays, then the problem of activated CCA subarray partition rises at the r-UAV side for the fast multi-UAV beam tracking. The dynamic CCA subarray partition can be considered as the dynamic antenna resource allocation for multiple t-UAVs, which has strong impact on the sum SE of the UAV mmWave network.

From the aforementioned two properties of the CCA, we know that the optimal beamforming and combining vector $\boldsymbol{f}_k(t)$ and $\boldsymbol{w}_k(t)$ are relevant to the activated subarray and subarray partition. To this end, we can continue to reformulate the beam tracking problem by taking the subarray activation/partition into account, which will facilitate the specialized codebook design in the next section.

%\emph{Remark}��It is worth pointing out that subarray-dependent

Let us denote $\mathcal{S}^{\text{t}}_{k}$ as the activated subarray at the $k$-th t-UAV with $\ensuremath{|\mathcal{S}_{k}^{\text{t}}|=M_{\text{act},t,k}N_{\text{act},t,k}\leq M_{t}}N_{t}$, $\mathcal{S}^{\text{r}}_{k}$ the activated subarray for the $k$-th t-UAV at the r-UAV, and the array partition at the r-UAV as $\mathcal{S}^{\text{r}}=\mathop{\cup}\limits_{{k\in\mathcal{K}}}{\mathcal{S}^{\text{r}}_k}$ where $|\mathcal{S}_{k}^{\text{r}}|>0$, ${\mathcal{S}_{k}^{\text{r}}}\mathop{\cap}{\mathcal{S}_{j}^{\text{r}}}=\emptyset,k\neq j$ and $\ensuremath{|\mathcal{S}^{\text{r}}|=M_{\text{act},r}N_{\text{act},r}\leq M_{r}}N_{r}$. The SE maximization problem for the codebook-based beam tracking with CCA can be formulated as
%\begin{eqnarray}
%\label{P1}
%&\mathop{\max}\limits_{\boldsymbol{f}_{{k}},{\mathcal{S}^{\text{t}}_k},\boldsymbol{w}_{{k}},{\mathcal{S}^{\text{r}}_k}} R(t)\\
%&\text{s.t.}\boldsymbol{f}_{{k}}\in\mathcal{F}\\ \nonumber
%&\ \ \ \boldsymbol{w}_{{k}}\in\mathcal{W}_k\\ \nonumber
%&{\mathcal{S}^{\text{r}}_k}\mathop{\cap}{\mathcal{S}^{\text{r}}_j}=\emptyset, k\neq j \nonumber
%\end{eqnarray}
%\begin{eqnarray}
%\begin{aligned}
%\label{P1}
%& \underset{\boldsymbol{f}_{{k}},{\mathcal{S}^{\text{t}}_k},\boldsymbol{w}_{{k}},{\mathcal{S}^{\text{r}}_k}}{\text{max}}
%& & R(t)\\
%& \text{subject to}
%& & \boldsymbol{f}_{{k}}\;\in\mathcal{F}, \\
%&&& \boldsymbol{w}_{{k}}\;\in\mathcal{W}_k,\\
%&&& {\mathcal{S}^{\text{r}}_k}\mathop{\cap}{\mathcal{S}^{\text{r}}_j}=\emptyset, k\neq j.
%\end{aligned}
%\end{eqnarray}%\end{equation*}
\begin{eqnarray}
\begin{aligned}\label{P1} & \underset{\boldsymbol{f}_{{k}},{\mathcal{S}_{k}^{\text{t}}},\boldsymbol{w}_{{k}},{\mathcal{S}_{k}^{\text{r}}}}{\text{max}} &  & R(t)\\
 & \text{subject to} &  & \boldsymbol{f}_{{k}}\left(\mathcal{S}_{k}^{\text{t}}\right)\;\in\mathcal{F},\\
 &  &  & \boldsymbol{w}_{{k}}\left(\mathcal{S}_{k}^{\text{r}}\right)\;\in\mathcal{W},\\
 &  &  & \ensuremath{\mathcal{S}^{\text{r}}=\mathop{\cup}\limits _{{k\in\mathcal{K}}}{\mathcal{S}_{k}^{\text{r}}}},\\
 &  &  & {\mathcal{S}_{k}^{\text{r}}}\mathop{\cap}{\mathcal{S}_{j}^{\text{r}}}=\emptyset,\thinspace k\neq j.
\end{aligned}
\end{eqnarray}
$\mathcal{F}$ and $\mathcal{W}$ are the sets of all analog beamforming vectors and combing vectors satisfying the hardware constraints, respectively.
In fact, solving the above problem (\ref{P1}) requires the new codebook design and codeword selection/processing strategy. Noting the interdependent relationship between the beamformer/combiner (or AWV) and the activated subarray or subarray partition, a well-structured codebook should be designed to facilitate the fast localization of the activated subarray and flexible beam control. For this goal, the CCA codebook design and the codebook-based joint Subarray Partition and Array-weighting-vector Selection (SPAS) algorithm will be first proposed in the next section.

In addition, the AOAs and AODs should be tracked in the highly dynamic UAV mmWave network.
To this end, in Section~\ref{Sec4} we will further propose a novel predictive AOA/AOD tracking scheme in conjunction with tracking error treatment to address the high mobility challenge, then we integrate these operations into the codebook-based SPAS to achieve reliable beam-tracking for the considered UAV mmWave network.
\section{CCA Codebook Design and Joint Subarray Partition and Code Selection}
%����������Ϊ������ �ŵ���Ϣ��ȡ �� ��֪AOA AOD �������ʱ����������ѡ��
\label{Sec3}
In this section, we characterize the CCA from several relevent aspects in \ref{Sec3.1} and design a specialized hierarchical codebook for the DRE-covered CCA in \ref{Sec3.2}, wherein the subarray activation/partitioning patterns (in terms of subarray location and size) are carefully integrated with the angular domain beam patterns (in terms of beam angles and widths). Then,
we present the basic framework of the codebook-based SPAS for the beam tracking in the considered UAV mmWave network for t-UAVs and the r-UAV in \ref{Sec3.3} and \ref{Sec3.4}, respectively.

\subsection{CCA Characterization}
\label{Sec3.1}
In this subsection, we characterize the CCA from serval relevant aspects to facilitate the codebook design.
\subsubsection{Maximum Resolution}
%The CCA codebook design is based on the characteristics of the DRE-covered CCA, which are further explained with more details here.
 The first characteristic is the maximum codebook resolution which is related to the maximum size of the activated subarray of the DRE-covered CCA. More specifically, due to the directivity of the antenna element, only a subset of the elements satisfying $\alpha_0+2l\pi\in[\alpha_{n,\min},\alpha_{n,\max}]$ can be activated at a certain beam angle $\alpha_0$, as shown in Fig.~\ref{figsim1}. Hence, the maximum number of elements which can be activated, i.e., the maximum resolution, is given by the following theorem.
\newtheorem{theorem}{Theorem}
\begin{theorem}
\label{the1}
For an $M \times N$-element DRE-covered CCA, the maximum number of the activated elements on the $xy$-plane with a given azimuth angle $\alpha_0$ is given by $N_{\text{act,max}}$.
\begin{eqnarray}
\label{maxN}
\begin{aligned}
&N_{\text{act,max}}=|n_2-n_1|,\\
&n_1=\biggl\lceil{\frac{\alpha_0+2l\pi-\Delta\alpha/2}{\Delta\phi}+\frac{N+1}{2}}\biggr\rceil,\\
&n_2=\biggl\lceil{\frac{\alpha_0+2l\pi+\Delta\alpha/2}{\Delta\phi}+\frac{N+1}{2}}\biggr\rceil,
\end{aligned}
\end{eqnarray}
%where $d_c(n_1,n_2)$ is the distance between element $n_1$ and $n_2$ on $xy$-plane.If $|n_{1}-n_{2}|>N/2$, $d(n_{1},n_{2})=N-|n_{1}-n_{2}|$;otherwise, $d(n_{1},n_{2})=|n_{1}-n_{2}|$.
and the maximum number of the activated elements in the $z$-axis is $M$.
\end{theorem}%$BW_{\text{a(e)}}=\min\{BW_{\text{a(e),array}},BW_{\text{a(e),element}}\}$ �ŵ�theorem�� ǰ��ֻ������
\begin{IEEEproof}
Please refer to Appendix \ref{appendix:A}.
\end{IEEEproof}

According to Theorem~\ref{the1}, we know that all elements along the $z$-axis can be activated at a certain elevation angle while only a part of the elements on the $xy$-plane can be activated at a certain azimuth angle.

%Given the maximum resolution of the codebook, we continue to discuss the characteristic of beamwidth with the CCA codebook. To start with, we introduce the beam gain of an AWV $\boldsymbol{v}$ along the angle $(\alpha,\beta)$ as
%\begin{eqnarray}
%\label{beamgain}
%G(\boldsymbol{v},\alpha,\beta)=\sqrt{N_{\text{act}}M_{\text{act}}}\boldsymbol{a}(N,M,\alpha,\beta)^H\boldsymbol{v},
%\end{eqnarray}
%where $N_{\text{act}}M_{\text{act}}$ is the number of elements of $\boldsymbol{v}$.
\textcolor{black}{\subsubsection{Multi-resolution and Beamwidth}
% To start with, we introduce the beam gain of an subarray-dependent $MN\times1$ AWV $\ensuremath{\boldsymbol{v}}\left(\mathcal{S}_{\mathrm{act}}\right)$ along the angle $(\alpha,\beta)$ as
%\begin{eqnarray}
%\label{beamgain}
%G(\boldsymbol{v},\alpha,\beta)=\sqrt{N_{\text{act}}M_{\text{act}}}\boldsymbol{a}(\mathcal{S}_{\mathrm{act}},\alpha,\beta)^H\ensuremath{\boldsymbol{v}}\left(\mathcal{S}_{\mathrm{act}}\right),
%\end{eqnarray}
%where $N_{\text{act}}M_{\text{act}}$ is the size of the activated subarray $\mathcal{S}_{\mathrm{act}}$ to support $\boldsymbol{v}$ and $\boldsymbol{a}(\mathcal{S}_{\mathrm{act}},\alpha,\beta)$ is the corresponding array steering vector along the angle $(\alpha,\beta)$. Then we define the azimuth beam coverage of $\boldsymbol{v}$ as
%\begin{eqnarray}
%\label{cva}
%\mathcal{CV}_{\text{a}}(\boldsymbol{v})=\{\alpha|G>\rho\mathop{\max}\limits_{\alpha^{*}}G(\boldsymbol{v},\alpha^{*},\beta_0)\},
%\end{eqnarray}
%where $\beta_0$ is a reference elevation angle, $\rho$ is the beam coverage factor within $(0,1)$. Similarly, the elevation beam coverage of $\boldsymbol{v}$ is defined as
%\begin{eqnarray}
%\label{cve}
%\mathcal{CV}_{\text{e}}(\boldsymbol{v})=\{\beta|G>\rho\mathop{\max}\limits_{\beta^{*}}G(\boldsymbol{v},\alpha_0,\beta^{*})\},
%\end{eqnarray}
%where $\alpha_0$ is a reference azimuth angle. When $\rho=\frac{1}{\sqrt{2}}$, the beam coverage is the 3dB beam coverage and the corresponding beamwidth is 3dB beamwidth or the Half-Power Beamwidth (HPBW).
Given the maximum resolution of the codebook, we continue to discuss the characteristic of the multi-resolution and the beamwidth with the CCA codebook. For the multi-resolution codebook, the variable resolution is tuned by the beamwidth, which is determined by the number of the activated elements~\cite{13}. Note that the beam coverage and the corresponding beamwidth are determined by both the element radiation pattern and array radiation pattern for the DRE-covered CCA. In particular, the beam coverage in the azimuth (elevation) plane of the activated $M_{\text{act}} \times N_{\text{act}}$ subarray is
\begin{eqnarray}
\label{CV}
\mathcal{CV}_{\text{a(e)}}=\mathcal{CV}_{\text{a(e),array}}\cap\mathcal{CV}_{\text{a(e),element}},
\end{eqnarray}
where $\mathcal{CV}_{\text{a(e),array}}$ is the coverage of the antenna array with omnidirectional elements~\cite{13}, named as the array coverage, and
$\mathcal{CV}_{\text{a(e),element}}=\{\alpha(\beta)|\lambda_s(\alpha,\beta)>0\}$ is the coverage of the $M_{\text{act}}N_{\text{act}}$ antenna elements, named as the element coverage. $\lambda_s(\alpha,\beta)$ is the sum antenna elements gain at the angle $(\alpha,\beta)$ that is given by
 $\lambda_s(\alpha,\beta)=\sum_{m,n}\left[{{\boldsymbol{\Lambda }}}\left( {{\alpha}},{\beta} \right) \right]_{(m,n)}$. The array coverage and the element coverage will be further explained in Section~\ref{Sec3.2} to analyze the coverage performance of the DRE covered CCA codebook.
The corresponding beamwidth of the azimuth (elevation) plane of the DRE-covered CCA $BW_{\text{a(e)}}$ is represented as
\begin{eqnarray}
\label{BW}
BW_{\text{a(e)}}=\min\{BW_{\text{a(e),array}},BW_{\text{a(e),element}}\},
\end{eqnarray}
where $BW_{\text{a(e),array}}$ is the beamwidth of the antenna array with omnidirectional elements, named as the array beamwidth, which is usually defined by the number of elements in the codebook design~\cite{13}. In the designed codebook, $BW_{\text{a(e),array}}$ is set as $BW_{\text{a,array}}=\frac{2\pi}{N_{\text{act}}}$ and $BW_{\text{a,array}}=\frac{2\pi}{M_{\text{act}}}$.
The element beamwidth $BW_{\text{a(e),element}}$ is the width of the element coverage $\mathcal{CV}_{\text{a(e),element}}$.
% $\lambda_s(\alpha,\beta)$ is the sum antenna elements gain at angle $(\alpha,\beta)$ that is given by
% $\lambda_s(\alpha,\beta)=\sum_{m,n}\left[{{\boldsymbol{\Lambda }}}\left( {{\alpha }},{\beta} \right) \right]_{(m,n)}$.
Note that for DRE-covered CCA, we still need the following theorem to determine $BW_{\text{a(e),element}}$ so as to give a full description of the beamwidth in (\ref{BW}).}
\begin{theorem}
\label{the2}
%The $M \times N$ CCA' array 3dB beamwidth of the azimuth plane and the elevation plane is $\frac{2\pi}{N}$ and $\frac{2\pi}{M}$, respectively.
\textcolor{black}{When $\Delta\phi_{\text{c}}\leq\Delta\alpha$,} the DRE coverage of $M \times N$ CCA on the azimuth plane is
\begin{eqnarray}
\label{elebw_az}
%BW_{\text{a,element}}=\alpha_{\text{max}}-\alpha_{\text{min}}+\left[\biggl\lfloor{(1-2\rho) N}\biggr\rfloor+1\right]\Delta\phi,
BW_{\text{a,element}}=\Delta\alpha+(N-1)\Delta\phi+2l\pi,l\in\mathbb{Z},
\end{eqnarray}
and the DRE coverage on the elevation plane is
\begin{eqnarray}
\label{elebw_ele}
BW_{\text{e,element}}=\Delta\beta.
\end{eqnarray}
\end{theorem}%$BW_{\text{a(e)}}=\min\{BW_{\text{a(e),array}},BW_{\text{a(e),element}}\}$ �ŵ�theorem�� ǰ��ֻ������
\begin{IEEEproof}
Please refer to Appendix \ref{appendix:B}.
\end{IEEEproof}

%According to the Theorem 2, the CCA's element beamwidth of azimuth plane is connected with the size of the activated CCA array. The reason is that different number of elements can be activated at
%different azimuth angles, which is in accord with the Theorem~1.
%When the omni-directional elements are used, the beamwidth is only determined by the array radiation pattern, i.e., $BW_{\text{a(e)}}=BW_{\text{a(e),array}}, \mathcal{CV}_{\text{a(e),element}}=2 \pi$.
\subsubsection{Localized Subarray Activation}
According to Theorem~\ref{the1}, only a subarray of CCA can be activated at a certain beam angle. Next, the relationship between the subarray and the beam angles is studied. The number and position of the activated elements determine the subarray. %The distance between the elements of antenna array have influence on the radiation pattern of the array.%(����������)
\textcolor{black}{Assuming that the elements in the activated subarray are adjacent to each other and the activated subarray can be expanded as a rectangle, the activated subarray is \textcolor{kkb}{denoted} by the set $\mathcal{S}(M_{\text{act}},N_{\text{act}},\boldsymbol{p}_c)$,} where $M_{\text{act}}$ and $N_{\text{act}}$ are the numbers of elements on the $z$-axis and $xy$-plane, respectively, and $\boldsymbol{p}_c=(m_c,n_c)$ is the position of the subarray center element, which is related to the azimuth angle. Then we have the following theorem to localize the activated subarray.
\begin{theorem}
\label{the3}
For the $M \times N$-element CCA, given the beam angle $(\alpha,\beta)$ and the size of subarray $(M_{\text{act}},N_{\text{act}})$, when the position of the center element of the subarray $\boldsymbol{p}_c$ satisfies
\begin{eqnarray}
\begin{aligned}
\label{ncm}
%&n_{c}^{\text{max}}=\mod\left(\biggl\lceil{\frac{\alpha}{\Delta\phi}+(N+1)/2}\biggr\rceil,N\right),\\
&n_{c}=\biggl\lceil{\frac{\alpha+2l\pi}{\Delta\phi}+(N+1)/2}\biggr\rceil, l\in\mathbb{Z},\\
&m_{c}=\biggl\lfloor{\frac{M}{2}}\biggr\rfloor,%\\
%\label{nc}
%&n_{c}\in\mathcal{D}_n=\left[\biggl\lceil{n_{c}^{\text{max}}-\Delta n}\biggr\rceil,\biggl\lfloor{n_{c}^{\text{max}}\Delta n}\biggr\rfloor\right],\\
%\label{mc}
%&m_c\in\mathcal{D}_m=\left[\biggl\lfloor{\frac{M_{\text{act}}}{2}}\biggr\rfloor,\biggl\lfloor{M-\frac{M_{\text{act}}}{2}}\biggr\rfloor\right],
\end{aligned}
\end{eqnarray}
all elements of the subarray can be activated.%where $n_{c}^{\text{max}}$ is the position of the center element on the $xy$-plane when $N_{\text{act}}=N_{\text{act,max}}$,  $m_{c}^{\text{max}}$ is the position of the center element on the $z$-axis when $M_{\text{act}}=M_{\text{act,max}}$.
%and $\Delta n=\frac{N_{\text{act,max}}-N_{\text{act}}+1}{2}$.
\end{theorem}%$BW_{\text{a(e)}}=\min\{BW_{\text{a(e),array}},BW_{\text{a(e),element}}\}$ �ŵ�theorem�� ǰ��ֻ������
\begin{IEEEproof}
Please refer to Appendix \ref{appendix:C}.
\end{IEEEproof}
Theorem 3 provides the feasible position of the activated subarray's center element, which is related to the azimuth angle $\alpha$. This property indicates that when using the DRE-covered CCA, the activated subarray should be localized with reference to the azimuth angle.

\begin{figure}[!t]
\centering
\includegraphics[scale=1,width=3.4in]{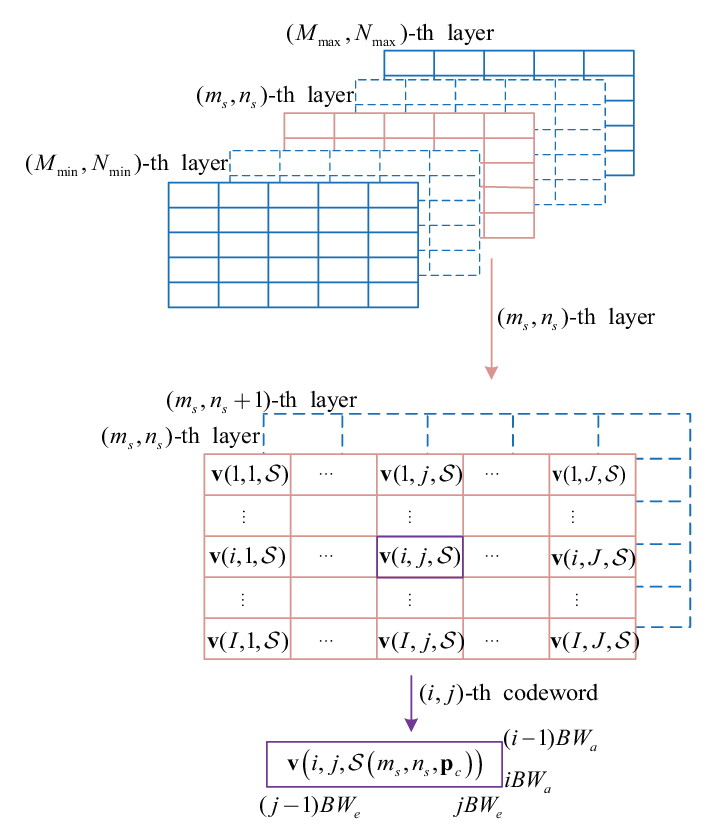}
\caption{The proposed 3D CCA hierarchical codebook. The hierarchical codebook contains multiple layers with different beamwidth. The $(i,j)$-th code of the $(m,n)$-th layer contains the AWV $\boldsymbol{v}$ and the corresponding subarray $\mathcal{S}$. The beam coverages of $\boldsymbol{v}(m_s,n_s,i,j,\mathcal{S})$ in azimuth angle and elevation angle are $[(i-1){BW}_{\text{a}},i{BW}_{\text{a}}]$ and $[(j-1){BW}_{\text{e}},j{BW}_{\text{e}}]$, respectively.}
\label{figsim3}
%\vspace{-15pt}
\end{figure}
\subsection{CCA Codebook Design}
\label{Sec3.2}
After the discussion on the characteristics of CCA, in this subsection, we continue to explain the specialized codebook design for the DRE-covered CCA. Revisiting Theorem~\ref{the1} and Theorem~\ref{the3}, the size and position of the activated CCA subarray are related to the azimuth angle; meanwhile, the beamwidth is determined by the size of the activated subarray according to Theorem~\ref{the2}. Therefore, the conventional codebook only consisting of different beamwidth and beam angles is not able to reveal the relationship among the beam angle, beamwidth and the corresponding supporting subarray for the DRE-covered CCA. In order to solve the beam tracking problem in (\ref{P1}), the subarray activation/partition and AWV selection needs to be jointly optimized at the same time. To this end, a new specialized hierarchical codebook $\mathcal{V}$ should be designed to facilitate efficient beam tracking, wherein the codeword $\boldsymbol{v}$ should contain both the angular-domain beam pattern information $(\alpha_i,\beta_i)$ and the corresponding subarray patten information $\mathcal{S}$.

As shown in Fig.~\ref{figsim3}, we propose a hierarchical codebook $\mathcal{V}$ that consists of multiple layers, and each layer is determined by the size of an activated subarray $(M_{\text{act}}=m_s,N_{\text{act}}=n_s)$. In a certain layer indexed by the subarray size $(m_s,n_s)$, the $(i,j)$-th codeword of AWV $\boldsymbol{v}(i,j,\mathcal{S})$ is determined by the supporting subarray $\mathcal{S}$, and the quantized azimuth angle $\alpha_i=i\frac{BW_{\text{a}}}{2}$ and the quantized elevation angle $\beta_j=j\frac{BW_{\text{e}}}{2}$, where $i\in\mathcal{I}_{n_s}=\left[1,I=\lceil\frac{2\pi}{BW_{\text{a}}}\rceil\right],j\in\mathcal{J}_{m_s}=\left[1,J=\lceil\frac{\pi}{BW_{\text{e}}}\rceil\right]$. In particular, the supporting subarray $\mathcal{S}(m_s,n_s,\boldsymbol{p}_c(i))$ has a size of $(m_s,n_s)$, and its center element location $\boldsymbol{p}_c(i)=(m_c,n_c(i))$ is relevant to the azimuth angle, i.e., the index $i$ of the azimuth angle. Since the beamwidth is mainly effected by the size of the subarray according to Theorem~\ref{the2}, the codewords of AWV in the same layer has the same beamwidth. Moreover, given the size of supporting array $(m_s,n_s)$ and the beam angle $(\alpha_i,\beta_j)$, the center element position of the supporting subarray $\boldsymbol{p}_c(i)=(m_c,n_c(i))$ is given by (\ref{ncm}) according to Theorem~\ref{the3}.%\emph{Remark 1}:
\textcolor{black}{
\begin{remark}
\label{R1}
%The multiple resolution codebooks usually focus on both the beam direction and the beamwidth.
The conventional UPA/ULA codebook design mainly controls the beamwidth by the subarray activation/deactivation with different numbers of elements. In contrast, the codebook for DRE-covered CCA focuses on both the number of subarray elements and the specific subarray localization. The number of subarray elements determines the beamwidth and the localization determines the beam angle that can be achieved by the subarray.
\end{remark}}

Along with the overall codebook structure, we continue to reveal the inner contents of the codeword $\boldsymbol{v}(i,j,\mathcal{S})$ in the $(m_s,n_s)$-th layer. Note that $\boldsymbol{v}(i,j,\mathcal{S})$ has a size of $MN$ and its $(m,n)$-th entry is given by
\begin{eqnarray}
\label{vcode1}
[\boldsymbol{v}(i,j,\mathcal{S})]_{(m,n)}=\boldsymbol{1}_{\{m,n,i,j\}}[\boldsymbol{A}(\alpha_i,\beta_j)]_{(m,n)},
\end{eqnarray}
where $\boldsymbol{A}(\alpha_i,\beta_j)$ is given in (\ref{steering}) and $\boldsymbol{1}_{\{m,n,i,j\}}$ is the indicator function
\begin{eqnarray}
\label{indictor}
\boldsymbol{1}_{\{m,n,i,j\}}=
\left\{ \begin{IEEEeqnarraybox}[\relax][c]{ll}
  & 1,(m,n)\in\mathcal{S}(m_s,n_s,\boldsymbol{p}_c(i)), \\
 &  0,\text{else}.
\end{IEEEeqnarraybox} \right.
\end{eqnarray}
\textcolor{black}{According to \eqref{vcode1}, the codeword $\boldsymbol{v}(i,j,\mathcal{S})$ includes both the beam pattern information and the subarray pattern information. The beam pattern information mainly \textcolor{kkb}{includes} the beam angle $(\alpha_i,\beta_j)$ and the beam width determined by the size of $\mathcal{S}$\textcolor{kkb}{;} the subarray pattern information \textcolor{kkb}{includes} the subarray location and size determined by $\mathcal{S}$.}

%The beam coverage of $v(m,n,i,j)$ is given by
%\begin{eqnarray}
%\label{cvbeam}
%&\mathcal{CV}_{\text{a}}(v(m,n,i,j))=\left[\frac{(i-1)BW_{\text{a}}}{2},\frac{(i+1)BW_{\text{a}}}{2}\right],\\
%&\mathcal{CV}_{\text{e}}(v(m,n,i,j))=\left[\frac{(j-1)BW_{\text{e}}}{2},\frac{(j+1)BW_{\text{e}}}{2}\right].
%\end{eqnarray}
%������ȫ���ǵ����� ˵��ʵ���г��õĶ���������������������
%When the element beamwidth on the elevation plane $BW_{\text{e,element}}=\pi$ and the element beamwidth on the azimuth plane $BW_{\text{e,element}}>\Delta\phi$, the beam coverage of all AWVs in the $(m,n)$-th layer is
%\begin{eqnarray}
%\label{cvlayera}
%&\mathcal{CV}_{\text{a}}(v(m,n,j))=\mathop{\cup}\limits_{i}\mathcal{CV}_{\text{a}}(v(m,n,i,j))=2\pi,\\
%\label{cvlayere}
%&\mathcal{CV}_{\text{e}}(v(m,n,i))=\mathop{\cup}\limits_{j}\mathcal{CV}_{\text{e}}(v(m,n,i,j))=\pi.
%\end{eqnarray}
%In this case, the CCA has the omnidirectional coverage.
Finally, we elaborate the coverage property of our codebook. The coverage of the $(i,j)$-th codeword in the $(m_s,n_s)$-th layer is given by
\begin{eqnarray}
\label{CV_code}
\mathcal{CV}_{\text{a(e)}}(i,j,\mathcal{S})=\mathcal{CV}_{\text{a(e),array}}(i,j,\mathcal{S})\cap\mathcal{CV}_{\text{a(e),element}}(i,j,\mathcal{S}).
\end{eqnarray}
\textcolor{black}{Then the coverage property is characterized by the following theorem.
\begin{theorem}
\label{the4}
 When the $(i,j)$-th codeword in each layer satisfies $i\in\mathcal{I}_{n_s}=\left[1,I=\lceil\frac{2\pi}{BW_{\text{a}}}\rceil\right]$ and $j\in\mathcal{J}_{m_s}=\left[1,J=\lceil\frac{\pi}{BW_{\text{e}}}\rceil\right]$,
the union of the beam coverage of all codewords in each layer covers the whole azimuth and elevation angular domain, i.e.,
\begin{eqnarray}
 \label{coveragepro}
\left\{ \begin{IEEEeqnarraybox}[\relax][c]{ll}
&\underset{i\in \mathcal{I}}{\cup}\mathcal{CV}_{\text{a}}(i,j,\mathcal{S})=[\alpha_{0},\alpha_{0}+2\pi],\\
&\underset{j\in \mathcal{J}}{\cup}\mathcal{CV}_{\text{e}}(i,j,\mathcal{S})=[\beta_{0},\beta_{0}+2\pi],
 \end{IEEEeqnarraybox} \right.
 \end{eqnarray}
 where $\alpha_{0}$ and $\beta_{0}$ is an arbitrary angle.
\end{theorem}%$BW_{\text{a(e)}}=\min\{BW_{\text{a(e),array}},BW_{\text{a(e),element}}\}$ �ŵ�theorem�� ǰ��ֻ������
\begin{IEEEproof}
Please refer to Appendix \ref{appendix:D}.
\end{IEEEproof}}

\begin{figure}[!t]
\centering
\includegraphics[scale=0.8,width=2.5in]{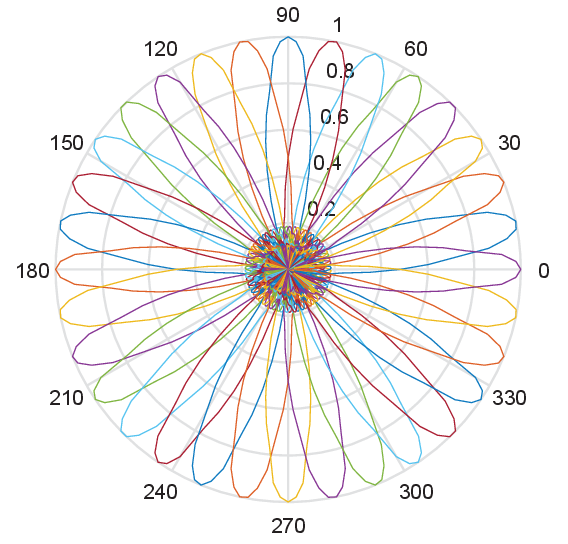}
\caption{Polar plots of the $(4,32)$-th layer of the codebook coverage on the azimuth plane with $BW_{\text{a,array}}=\frac{\pi}{16}$, $\Delta\alpha=\frac{2\pi}{3}$ and $\Delta\beta=\pi$.}
\label{pattern}
\end{figure}

In Fig.~\ref{pattern}, we show the polar plots of the $(4,32)$-th layer of the codebook coverage on the azimuth plane with $BW_{\text{a,array}}=\frac{\pi}{16}$, $\Delta\alpha=\frac{2\pi}{3}$ and $\Delta\beta=\pi$. It is observed that the AWVs of the same layer in the designed CCA codebook have the omnidirectional coverage.
%\rhoȡ���� ��ӦBWȡ���� �������� �뱾ʾ��ͼ ����ͼ�ķ�����������ͬ���ĸ���������
\subsection{CCA Codebook Based Joint Subarray Partition and AWV Selection Algorithm for t-UAVs}
\label{Sec3.3}
%In order to ensure flight safety, the UAVs are usually separated by a certain distance. Moreover, the narrow beam is used in the CCA-enabled UAV mmWave communications to compensate the propagation loss. Therefore, when the number of UAVs is not large and the transmit power of UAVs is limited, the inter-UAV interference is not the main factor affecting the sum SE and can be neglected in the analog beamforming, which will be demonstrated by the simulation results in Section~\ref{Sec6}.
Based on the designed CCA codebook, the joint subarray partition and AWV selection (SPAS) algorithm is developed in this section to solve the beam tracking problem in (\ref{P1}).
As mentioned above, it is assumed that the inter-UAV interference approximately does not exist in the considered UAV mmWave networks and the sum SE in problem (\ref{P1}) is rewritten as
\begin{eqnarray}
\label{SE2}
R(t)=\sum_{k\in \mathcal{K}}\log(1+\text{SNR}_k(t)),
\end{eqnarray}
where $\text{SNR}_k(t)$ is given by
\begin{eqnarray}
\label{SNR}
\text{SNR}_k(t)\!=\!\frac{\left|\sqrt{p_{k}(t)}\boldsymbol{w}_{k}(t)^H\boldsymbol{H}_{k}(t)\boldsymbol{f}_{k}(t)\right|^2}{\!\boldsymbol{n}_{\sigma}(t)}.
\end{eqnarray}
Given the transmit power, the AOAs $\left\{(\alpha_{k}^{\text{r}}(t),\beta_{k}^{\text{r}}(t)) \right\}$ and AODs $\left\{(\alpha_{k}^{\text{t}}(t),\beta_{k}^{\text{t}}(t))\right\}$ at slot $t$, the beam tracking problem in (\ref{P1}) can be translated to maximize the beam gain of the beamforming vector $\boldsymbol{f}_{{k}}({\mathcal{S}_{k}^{\text{t}}})$ and the combining vector $\boldsymbol{w}_{{k}}({\mathcal{S}_{k}^{\text{r}}})$, respectively, without considering the interference between the t-UAVs. Then the problem can be solved by the SPAS algorithm for t-UAV and r-UAV, respectively.
To start with, the beam gain of a subarray-dependent $MN\times1$ AWV $\ensuremath{\boldsymbol{v}}\left(\mathcal{S}_{\mathrm{act}}\right)$ along the angle $(\alpha,\beta)$ is expressed as
\begin{eqnarray}
\label{beamgain}
G(\boldsymbol{v},\alpha,\beta)=\sqrt{N_{\text{act}}M_{\text{act}}}\boldsymbol{a}(\alpha,\beta, \mathcal{S}_{\mathrm{act}})^H\ensuremath{\boldsymbol{v}}\left(\mathcal{S}_{\mathrm{act}}\right),
\end{eqnarray}
where $N_{\text{act}}M_{\text{act}}$ is the size of the activated subarray $\mathcal{S}_{\mathrm{act}}$ to support $\boldsymbol{v}$ and $\boldsymbol{a}(\alpha,\beta, \mathcal{S}_{\mathrm{act}})$ is the corresponding array steering vector along the angle $(\alpha,\beta)$. The $(m,n)$-th entry of $\boldsymbol{a}(\alpha,\beta, \mathcal{S}_{\mathrm{act}})$ is expressed as
\begin{eqnarray}
\label{vcode}
[\boldsymbol{a}(\alpha,\beta, \mathcal{S}_{\mathrm{act}})]_{(m,n)}=\boldsymbol{1}_{\{\mathcal {S}_{\mathrm{act}}\}}[\boldsymbol{A}(\alpha,\beta)]_{(m,n)},
\end{eqnarray}
where $\boldsymbol{1}_{{\{\mathcal {S}_{\mathrm{act}}\}}}$ is the indicator function
\begin{eqnarray}
\label{indictor}
\boldsymbol{1}_{\{\mathcal{S}_{\mathrm{act}}\}}=
\left\{ \begin{IEEEeqnarraybox}[\relax][c]{ll}
  & 1,(m,n)\in \mathcal{S}_{\mathrm{act}}, \\
 &  0,\mathrm{else}.
\end{IEEEeqnarraybox} \right.
\end{eqnarray}

\begin{figure}[!t]
\centering
\includegraphics[scale=0.8, width=3.4in]{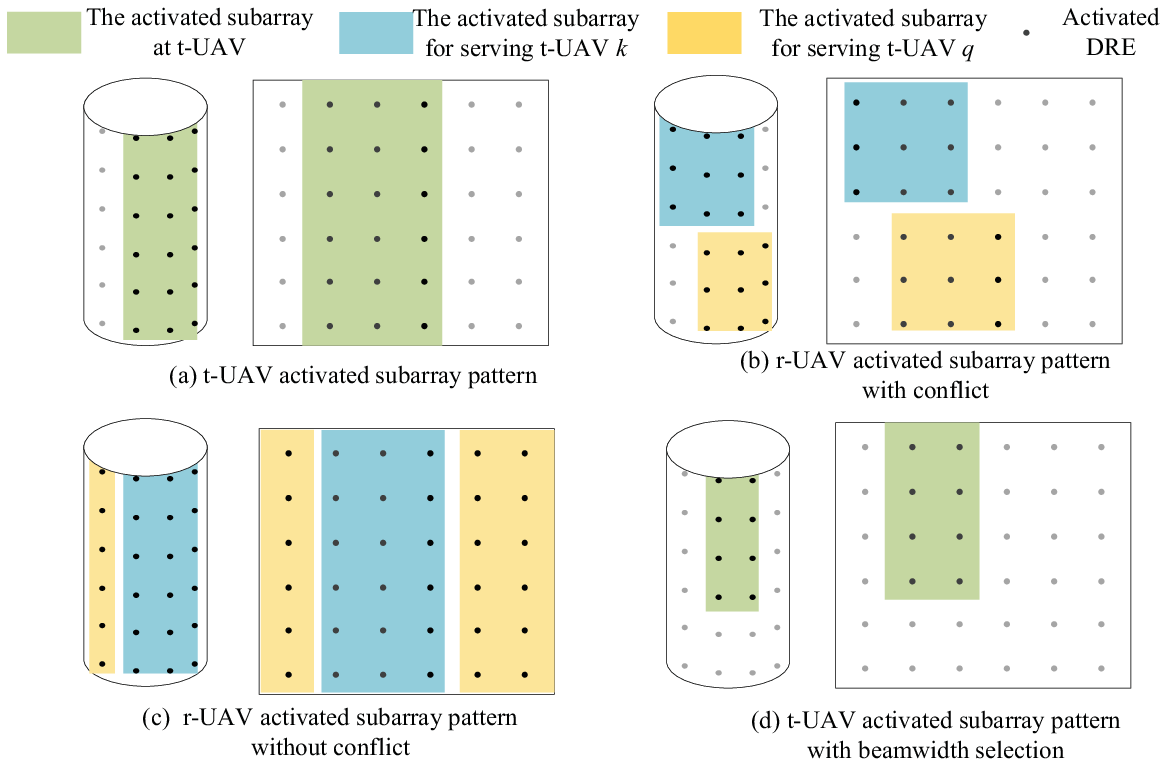}
\caption{The subarray patterns on the cylinder and the corresponding expanded cylinder. (a)~The t-UAV subarray partition pattern. (b)~The r-UAV subarray partition pattern with conflict. (c)~The r-UAV subarray partition pattern without conflict. (d)~The t-UAV subarray partition pattern with beamwidth selection.}
\label{sapattern}
\end{figure}

In the considered UAV mmWave network, the $k$-th t-UAV transmits to only one r-UAV. Hence,
%assuming a fixed combing vector $\boldsymbol{w}_k({\mathcal{S}_{k}^{\text{r}}})$ (and the activated subarrays ${\mathcal{S}_{k}^{\text{r}}}$),
the beam tracking problem for t-UAVs in (\ref{P1}) with our proposed codebook can be rewritten as
\begin{eqnarray}
\begin{aligned}\label{P2} & \underset{\boldsymbol{f}_{{k}},{\mathcal{S}_{k}^{\text{t}}}}{\text{max}} &  & G(\boldsymbol{f}_{k}\left(\mathcal{S}_{k}^{\text{t}}\right),\alpha_{t,k}(t),\beta_{t,k}(t))\\
%{\left|\boldsymbol{w}_{k}({\mathcal{S}_{k}^{\text{r}}};t)^{H}\boldsymbol{H}_{k}(t)\boldsymbol{f}_{k}({\mathcal{S}_{k}^{\text{t}}};t)\right|}^{2}\left|\boldsymbol{w}_{k}({\mathcal{S}_{k}^{\text{r}}})\right.
 & \text{subject to} &  & \boldsymbol{f}_{{k}}\left(\mathcal{S}_{k}^{\text{t}}\right)=\boldsymbol{v}(i,j,\mathcal{S})\;\in\mathcal{V}_{k}.
\end{aligned}
\end{eqnarray}
The t-UAV needs to select an appropriate codeword $\boldsymbol{v}(i,j,\mathcal{S})$ from our proposed codebook $\mathcal{V}_k$ to solve the subarray partition and AWV selection problem in (\ref{P2}). \textcolor{black}{Note that after the codeword $\boldsymbol{v}(i,j,\mathcal{S})$ is selected, the beam pattern and the subarray pattern \textcolor{kkb}{are} determined.}
%Let $\boldsymbol{f}_{{k}}\left(\mathcal{S}_{k}^{\text{t}}\right)=\boldsymbol{v}\left(m_s,n_s,i,j,\mathcal{S}\right)$ and $\mathcal{S}^{t}_k=\mathcal{S}(m_s,n_s,\boldsymbol{p}_{c})$, the problem in (\ref{P2}) can be further rewritten as the codebook based beam tracking problem
%\begin{eqnarray}
%\begin{aligned}\label{P3} & \underset{\boldsymbol{v}}{\text{max}} &  & \log(1+\gamma_k(t))\\
% & \text{subject to} &  & \boldsymbol{v}(m_s,n_s,i,j,\mathcal{S})\;\in\mathcal{V}.
%\end{aligned}
%\end{eqnarray}
 Given AODs, the maximum size of the activated subarray should be selected and the quantization error between the AODs and the beam angles in the codeword should be minimized to maximize the beam gain of the beamforming vector of the $k$-th t-UAV. Therefore, the optimal codeword $\boldsymbol{v}\left(i_{k}^{*},j_{k}^{*},\mathcal{S}\left(m_{s,k}^{*},n_{s,k}^{*},\boldsymbol{p}_{c,k}\left(i_{k}^{*}\right)\right)\right)$
 to solve the problem in (\ref{P2}) at slot $t$ is given by
\begin{eqnarray}
 \label{optv}
\left\{ \begin{IEEEeqnarraybox}[\relax][c]{ll}
 m_{s,k}^{*}&=M_{\text{act,max}},\\
 n_{s,k}^{*}&=N_{\text{act,max}},\\
 i_k^{*}&=\mathop{\text{arg min}}\limits_{i\in\mathcal{I}_{n_s}} |\alpha_{t,k}(t)-\alpha_i|,\\
 j_k^{*}&=\mathop{\text{arg min}}\limits_{j\in\mathcal{J}_{m_s}} |\beta_{t,k}(t)-\beta_j|,\\
 m_{c,k}^{*}&=\biggl\lfloor{\frac{M}{2}}\biggr\rfloor,\\
 n_{c,k}^{*}&=\biggl\lceil{\frac{\alpha_{i_{k}^{*}}+2l\pi}{\Delta\phi}+(N+1)/2}\biggr\rceil, l\in\mathbb{Z},
 \end{IEEEeqnarraybox} \right.
 \end{eqnarray}
 and the corresponding supporting subarray $\mathcal{S}(m_{s,k}^{*},n_{s,k}^{*},\boldsymbol{p}_{c,k}^{*})$ is selected as the activated subarray at the same time, here $\boldsymbol{p}^{*}_{c,k}=(m_{c,k}^{*}, n_{c,k}^{*})$, which is relevant to $i_k^{*}$. As shown in Fig.~\ref{sapattern} (a), the t-UAV subarray pattern is a part of cylinder with the size of $M_{\text{act}}=M_{\text{act,max}}$ and $N_{\text{act}}=N_{\text{act,max}}$.
\textcolor{black}{As $\alpha_i$ and $\beta_j$ is the quantization of azimuth angle and elevation angle, respectively, the indexes of the optimal codewords $i_k^{*}$ and $j_k^{*}$ in the given layer of the codebook according to \eqref{optv} are given by
           $i_k^{*}=\biggl\lceil{\frac{\alpha_{t,k}(t)}{BW_{a}}}\biggr\rceil$, and $j_k^{*}=\biggl\lceil{\frac{\beta_{t,k}(t)}{BW_{e}}}\biggr\rceil$.}
%If $\alpha_{t,k}(t)-\biggl\lfloor{\frac{\alpha_{t,k}(t)}{BW_{\text{a}}}}\biggr\rfloor\leq\frac{BW_{\text{a}}}{2}$,
%$i_k^{*}=\biggl\lfloor{\frac{\alpha_{t,k}(t)}{BW_{a}}}\biggr\rfloor$; if else,
%If $\beta_{t,k}(t)-\biggl\lfloor{\frac{\beta_{t,k}(t)}{BW_{\text{e}}}}\biggr\rfloor\leq\frac{BW_{\text{e}}}{2}$, %$j_k^{*}=\biggl\lfloor{\frac{\beta_{t,k}(t)}{BW_{e}}}\biggr\rfloor$; if else,

%\begin{eqnarray}
 %\label{opcdid}
 %i_k^{*}=\left\{ \begin{IEEEeqnarraybox}[\relax][c]{ll}
 %\biggl\lfloor{\frac{\alpha_{t,k}(t)}{BW_{a}}}\biggr\rfloor, \\
 %\biggl\lceil{\frac{\alpha_{t,k}(t)}{BW_{a}}}\biggr\rceil, else
 %\end{IEEEeqnarraybox} \right.
%\end{eqnarray}

% \textcolor{black}{\emph{Remark}: The different layers in the multiple resolution codebooks correspond to the different beamwidth, which also correspond to the different resolution. The beamwidth (resolution) is determined by the number of subarray elements, which is also called the size of the subarray.}
 %When the AOAs are estimated, the optimal solution $m^{*},n^{*},m_c^{*},n_c^{*}$ is effected by the estimation errors, which will be illustrated in Section \uppercase\expandafter{\romannumeral4}.
\subsection{CCA Codebook Based Joint Subarray Partition and AWV Selection Algorithm for r-UAV}
\label{Sec3.4}
In the considered UAV mmWave network, the r-UAV needs to activate multiple subarrays and select multiple combining vectors to serve multiple t-UAVs at the same time. Hence, the beam gain of the combining vector maximization problem for r-UAV with our proposed codebook can be rewritten as
\begin{eqnarray}
\begin{aligned}\label{P4} & \underset{\boldsymbol{w}_{{k}},{\mathcal{S}_{k}^{\text{r}}}}{\text{max}} &  &
G(\boldsymbol{w}_k\left(\mathcal{S}_{k}^{\text{r}}\right),\alpha_{r,k}(t),\beta_{r,k}(t))\\
%{\left|\boldsymbol{w}_{k}({\mathcal{S}_{k}^{\text{r}}};t)^H\boldsymbol{H}_{k}(t)\boldsymbol{f}_{k}({\mathcal{S}_{k}^{\text{t}}};t)\right|}^2\left|\boldsymbol{f}_k({\mathcal{S}_{k}^{\text{t}}})\right.\\
 & \text{subject to} &  &  \boldsymbol{w}_{{k}}\left(\mathcal{S}_{k}^{\text{r}}\right)=\boldsymbol{v}(i,j,\mathcal{S})\;\in \mathcal{V}_k, \\
 &  &  & \ensuremath{\mathcal{S}^{\text{r}}=\mathop{\cup}\limits _{{k\in\mathcal{K}}}{\mathcal{S}_{k}^{\text{r}}}},\\
 &  &  & {\mathcal{S}_{k}^{\text{r}}}\mathop{\cap}{\mathcal{S}_{j}^{\text{r}}}=\emptyset,\thinspace k\neq j.
\end{aligned}
\end{eqnarray}
The r-UAV needs to select multiple appropriate AWVs $\boldsymbol{v}(m_{s,k},n_{s,k},i_k,j_k,\mathcal{S}_k),k\in \mathcal{K}$ from our proposed codebook $\mathcal{V}$ to solve the subarray partition and AWVs selection problem. If an element is contained in different subarrays, there is a conflict between the subarrays. To solve the problem in (\ref{P4}), the joint SPAS problem without considering the conflict is discussed first and the conflict avoidance will be discussed later. Given AOAs, the maximum size of the activated subarray should be selected and the quantization error between the AOAs and the beam angles in the codeword should be minimized to maximize the beam gain of the combining vector for the $k$-th t-UAV.
%Let $\boldsymbol{w}_{{k}}=\boldsymbol{v}(m_{s,k},n_{s,k},i_k,j_k)$ and $\mathcal{S}^{r}_k=\mathcal{S}(m_{s,k},n_{s,k},\boldsymbol{p}_{c,k})$, the above problem can be further rewritten as
%\begin{eqnarray}
%\begin{aligned}\label{P5} & \underset{\boldsymbol{v}_k}{\text{max}} &  & R(t)\\
% & \text{subject to} &  & \boldsymbol{v}(m_{s,k},n_{s,k},i_k,j_k,\mathcal{S}_k)\;\in\mathcal{V}.
%\end{aligned}
%\end{eqnarray}
Similarly with (\ref{optv}),
the optimal codewords $\boldsymbol{v}\left(i_{k}^{*},j_{k}^{*},\mathcal{S}\left(m_{s,k}^{*},n_{s,k}^{*},\boldsymbol{p}_{c,k}\left(i_{k}^{*}\right)\right)\right)$ without considering the conflict constraint is expressed as
%\begin{eqnarray}
%\label{solutionr}
%\left\{ \begin{IEEEeqnarraybox}[\relax][c]{ll}
%m_{s,k}^{*}&=M_{\text{act,max}},\\
%n_{s,k}^{*}&=N_{\text{act,max}},\\
%i_{k}^{*}&=\mathop{\text{arg min}}\limits_{i\in \mathcal{I}_n} |\alpha_{r,k}(t)-\alpha_i|,\\
%j_{k}^{*}&=\mathop{\text{arg min}}\limits_{j \in \mathcal{J}_m} |\beta_{r,k}(t)-\beta_j|,\\
%m_{c,k}^{*}&=\biggl\lfloor{\frac{M}{2}}\biggr\rfloor,\\
%n_{c,k}^{*}&=\biggl\lceil{\frac{\alpha_{r,k}(t)+2l\pi}{\Delta\phi}+(N+1)/2}\biggr\rceil, l\in\mathbb{Z},
%\end{IEEEeqnarraybox} \right.
%\end{eqnarray}
\begin{eqnarray}
\label{solutionr}
\left\{ \begin{IEEEeqnarraybox}[\relax][c]{ll}
m_{s,k}^{*}&=M_{\text{act,max}},\\
n_{s,k}^{*}&=N_{\text{act,max}},\\
i_{k}^{*}&=\mathop{\text{arg min}}\limits_{i\in \mathcal{I}_n} |\alpha_{r,k}(t)-\alpha_i|,\\
j_{k}^{*}&=\mathop{\text{arg min}}\limits_{j \in \mathcal{J}_m} |\beta_{r,k}(t)-\beta_j|,\\
m_{c,k}^{*}&=\biggl\lfloor{\frac{M}{2}}\biggr\rfloor,\\
n_{c,k}^{*}&=\biggl\lceil{\frac{\alpha_{r,k}(t)+2l\pi}{\Delta\phi}+(N+1)/2}\biggr\rceil, l\in\mathbb{Z},
\end{IEEEeqnarraybox} \right.
\end{eqnarray}
and the activated subarray is obtained by the corresponding supporting subarray $\mathcal{S}(m_{s,k}^{*},n_{s,k}^{*},\boldsymbol{p}_{c,k}^{*})$, $\boldsymbol{p}_{c,k}^{*}=(m_{c,k}^{*},n_{c,k}^{*})$.
Compared to (\ref{optv}), $M_{\text{act,max}}$ and $N_{\text{act,max}}$ in~(\ref{solutionr}) are determined by the array size of r-UAV, $i_{k}^{*}$, $j_{k}^{*}$ and $n_{c,k}^{*}$ are determined by the AOAs $\left (\alpha_{r,k}(t),\beta_{r,k}(t)\right)$ instead of AODs. \textcolor{black}{The indexes of the optimal codeword $i_{k}^{*}$ and $j_{k}^{*}$ can be found in the way similar to \eqref{optv}. }

If there is no conflict between the activated subarrays, the subarray partition pattern is shown in Fig.~\ref{sapattern} (c).
However, it is possible that the optimal codewords for different t-UAVs need to activate the same antenna elements, which causes a conflict between the corresponding subarrays. Hence, the activated subarrays $\mathcal{S}_k^{\text{r}}$ should be partitioned based on the optimal codewords to avoid the conflict. To this end, the criterion to detect the conflict is discussed at first.
More specifically, if and only if
\begin{eqnarray}
\label{condition}
\begin{aligned}
d(n_{c,k},n_{c,q})<\frac{n_{s,k}+n_{s,q}}{2},\\
|m_{c,k}-m_{c,q}|<\frac{m_{s,k}+m_{s,q}}{2},%M<m_{s,k}+m_{s,q},
\end{aligned}
\end{eqnarray}
there is a conflict between the subarrays of $M\times N$-element CCA $\mathcal{S}_k(m_{s,k},n_{s,k}, \boldsymbol{p}_{c,k})$ and $\mathcal{S}_q (m_{s,q},n_{s,q},\boldsymbol{p}_{c,q})$, $(q\neq k)$. $d(n_{c,k},n_{c,q})$ is the distance between the center elements of the two subarrays on the $xy$-plane, which is given by
\begin{eqnarray}
\label{d_distance}
d(n_{c,k},n_{c,q})=
\left\{ \begin{IEEEeqnarraybox}[\relax][c]{ll}
N-|n_{c,k}-n_{c,q}|,|n_{c,k}-n_{c,q}|>N/2\\
|n_{c,k}-n_{c,q}|,\text{else},
\end{IEEEeqnarraybox} \right.
\end{eqnarray}
%If $|n_{c,k}-n_{c,q}|>N/2$, $d(n_{c,k},n_{c,q})=N-|n_{c,k}-n_{c,q}|$; otherwise, $d(n_{c,k},n_{c,q})=|n_{c,k}-n_{c,q}|$.
If (\ref{condition}) holds, there are not enough DREs between them to be activated separately and the conflict appears.

Then the conflict between the activated subarrays of the selected optimal codewords is detected according to (\ref{condition}). Denote $K\times K$ matrix $\boldsymbol{C}_{\text{sa}}$ as the conflict matrix, whose $(k,q)$-th element represents the conflict information between the subarray $\mathcal{S}_k$ and $\mathcal{S}_q$. If there is a conflict between the subarrays $\mathcal{S}_k$ and $\mathcal{S}_q$, $\boldsymbol{C}_{\text{sa},(k,q)}=1$, $q\neq k$; otherwise, $\boldsymbol{C}_{\text{sa},(k,q)}=0$. In addition, $\boldsymbol{C}_{\text{sa},(k,k)}=0$. In addition, if $\boldsymbol{C}_{\text{sa},(k,q)}=1$ and $\boldsymbol{C}_{\text{sa},(p,q)}=1$, $\boldsymbol{C}_{\text{sa},(k,p)}=1$. At last, the combining vector $\boldsymbol{w}_{{k}}=\boldsymbol{v}_k$ and the corresponding subarray $\mathcal{S}^{r}_k=\mathcal{S}_k$ needs to be updated to avoid the conflict.
The conflict set of the $k$-th subarray is defined as $\mathcal{C}_{\text{sa},k}=k\cup\{q|\boldsymbol{C}_{\text{sa},(k,q)}=1\}$.
$r_q$ is the index of the sorted elements of the conflict set $q$. The size of the activated subarray to serve the $q$-th UAV in the conflict set is updated by
\begin{eqnarray}
\label{updateM}
\left\{ \begin{IEEEeqnarraybox}[\relax][c]{ll}
N_{\text{act},q}&=n_{s,q},\\
M_{\text{act},q}&=\biggl\lfloor\frac{M}{|\mathcal{C}_{\text{sa},q}|}\biggr\rfloor.
\end{IEEEeqnarraybox} \right.
\end{eqnarray}
The combining vector to serve the $q$-th t-UAV in the conflict set is updated by
\begin{eqnarray}
\label{updatew}
[\boldsymbol{w}_q]_{(m,n)}=\left\{ \begin{IEEEeqnarraybox}[\relax][c]{ll}
&[\boldsymbol{w}_q]_{(m,n)}, m \in [(r_q-1)M_{\text{act},k}+1,r_qM_{\text{act},k}],\\
&0, \text{else}.
\end{IEEEeqnarraybox} \right.
\end{eqnarray}
The corresponding activated subarray is updated by
\begin{eqnarray}
\label{updatem}
\left\{ \begin{IEEEeqnarraybox}[\relax][c]{ll}
n_{c,q}&=n_{c,q}^{*},\\
m_{c,q}&=\biggl\lceil\frac{(2r_q-1)M_{\text{act}}+1}{2}\biggr\rceil.
\end{IEEEeqnarraybox} \right.
\end{eqnarray}
The conflict matrix $\boldsymbol{C}_{\text{sa}}$ is calculated again with the new activated subarrays. The updating procedure is finished if $\boldsymbol{C}_{\text{sa}}=\boldsymbol{0}$. Meanwhile, the subarray partition and AWV selection problem for the r-UAV is sloved. The corresponding subarray partition pattern is shown in Fig.~\ref{sapattern} (b).
%To mitigate the interference, if $k>1$, we romove component of the previous determined code from the $k$-th code by Gram-Schmidt proceduce:
%\begin{eqnarray}
%\label{GS}
%\boldsymbol{v}_k=\boldsymbol{v}_k-\sum\limits_{q=1}^{k-1}{\boldsymbol{v}_q^{H}\boldsymbol{v}_k\boldsymbol{v}_q}, \boldsymbol{v}_k=\boldsymbol{v}_k/||\boldsymbol{v}_k||,
%\end{eqnarray}

To summarize, the proposed CCA codebook based SPAS algorithm for r-UAV is given in Algorithm~\ref{AL1}.

\textcolor{black}{\emph{Complexity Analysis}: For each slot, the complexity of obtaining the optimal codewords is $\mathcal{O}(K)$; the complexity of calculating the conflict matrix and the conflict set is $\mathcal{O}(K^3)$; the combining vector and the activated subarray update takes $\mathcal{O}(M_{\text{act}}N_{\text{act}})$ operations. \textcolor{kkb}{In the worst case, there are conflicts among all the optimal activated subarrays, which causes the updating procedure's complexity of $\mathcal{O}(K^3+KM_{\text{act}}N_{\text{act}})$. Finally, the complexity of Algorithm~1 is on the order of \textcolor{kkb}{$\mathcal{O}(K^3+KM_{\text{act}}N_{\text{act}})$}. In the ideal case, there is no conflict among all the subarrays, and the updating procedure is not performed. In this case, the complexity of Algorithm~1 is $\mathcal{O}(K^3)$.}}

\begin{algorithm}[h]
\caption{CCA Codebook Based Joint Subarray Partition and Code Selection Algorithm for r-UAV}
\label{AL1}
\begin{algorithmic}[1]
\INPUT $(\alpha_{r,k}(t),\beta_{r,k}(t))$,$\mathcal{V}$, $\mathcal{K}$
\STATE Obtain the optimal codewords for $k\in\mathcal{K}$ according to (\ref{solutionr}).
\STATE Calculate the conflict matrix $\boldsymbol{C}_{\text{sa}}$ according to (\ref{condition}).
\WHILE {$\boldsymbol{C}_{\text{sa}}\neq \boldsymbol{0}, k \in\mathcal{K}$}
\STATE Calculate the conflict set of $k$-th subarray $\mathcal{C}_{\text{sa},k}$.
\FOR {$q \in \mathcal{C}_{\text{sa},k}$}
\STATE Update the combining vector $\boldsymbol{w}_q$ and the activated subarray $\mathcal{S}^{r}_q$ according to \eqref{updateM}, \eqref{updatew} and \eqref{updatem}.
\ENDFOR
\ENDWHILE
\OUTPUT $\boldsymbol{v}_k,\mathcal{S}^{r}_k$.
\end{algorithmic}
\end{algorithm}

\section{Tracking-Error-Aware Beam Tracking by Exploiting 3D Beamwidth Selection}
\label{Sec4}
The CCA codebook based SPAS algorithm is proposed in the previous section to solve the joint CCA subarray partition and AWV selection problem. In this section, the TE-aware beam tracking problem is addressed \textcolor{black}{based on the CCA codebook based SPAS algorithm}.
%The beam tracking problem in the considered UAV mmWave networks can be decomposed into two subproblems. One is the AOAs and AODs tracking problem and the other is the CCA joint subarray partition and code seleciton problem.
%The former is discussed given the perfect AOAs and AODs in Section \uppercase\expandafter{\romannumeral3}. In this section, the AOAs and AODs tracking problem is addressed.
Tracking the AOAs and AODs is essential for beam tracking, which is closely connected with the position and attitude of the t-UAVs and r-UAV. The position and attitude compose the UAV's motion state information (MSI). In this section, the MSI prediction based AOAs and AODs estimation scheme and the protocol for beam tracking are introduced in Section \ref{secIV.A}. Then the TE estimation algorithm which exploits the MSI prediction error is proposed in Section \ref{secIV.B}. %The tracking error and the beamwidth are related to the beam gain and the SE of the UAV networks. Hence,
The TE-aware CCA codebook based 3D beamwidth selection algorithm is developed based on the TE estimation to achieve effective beam tracking in Section \ref{secIV.C}.
\subsection{UAV Motion State Information}
\label{secIV.A}
The AOAs and AODs of the LOS channel in (\ref{channel}) are mainly determined by the position and attitude of the t-UAVs and r-UAV. The t-UAV and r-UAV motion state information (MSI) mainly consists of the t-UAV and r-UAV's position and attitude, denoted as ${{\boldsymbol{X}}_{t(r)}}=\left( {{x}_{t(r)}},{{y}_{t(r)}},{{z}_{t(r)}} \right)$ and ${{\boldsymbol{\Theta }}_{t(r)}}=\left( {{\psi }_{t(r)}},{{\theta }_{t(r)}},{{\phi }_{t(r)}} \right)$, respectively. The velocity and acceleration vectors are given by $\boldsymbol{v}_{t(r)}(t)=\frac{\boldsymbol{X}_{t(r)}(t)-\boldsymbol{X}_{t(r)}(t-1)}{\delta{t}}$ and $\boldsymbol{a}_{t(r)}(t)=\frac{\boldsymbol{v}_{t(r)}(t)-\boldsymbol{v}_{t(r)}(t-1)}{\delta{t}}$, where $\delta{t}$ is the time slot duration. The position and attitude of CCA related to the UAV is denoted as $\boldsymbol{X}_{\text{CCA}}$ and $\boldsymbol{\Theta}_{\text{CCA}}$.

The UAVs' trajectory on the $xy$-plane is assumed to follow the Smooth-Turn mobility model~\cite{Smooth-Turn} that can capture the mobility of UAVs in the scenarios like patrolling. In this model, the UAV circles around a certain point on the horizontal plane (xy-plane) for an exponentially distributed duration until the UAV selects a new center point with the turning radius whose reciprocal obeys the normal distribution $\mathcal{N}(0,\sigma^2_r)$. According to~\cite{Smooth-Turn}, $\sigma^2_r$ plays an important role in the degree of randomness. The UAVs are in the state of uniform linear motion in the vertical direction with different velocity $v_{t(r),z}$, where $v_{t(r),z}$ obeys the uniform distribution $v_{t(r),z}\sim\mathcal{U}(v_{t(r),z,\text{min}},v_{t(r)z,\text{max}})$. Moreover, aiming to maintain the communication link with the r-UAV, the t-UAVs keep their positions in a limited region at arbitrary time where the distance between the t-UAV and the r-UAV is less than $D_{\text{r,max}}$. \textcolor{black}{The distance between UAVs is also limited no less than $D_{\text{r,min}}$ to ensure the flight safety.} The relationship between the position and attitude (equations (8)-(10) in~\cite{attitude}) is used to determine the UAVs' attitude.

\begin{figure}[!t]
\centering
\includegraphics[height=0.9in,width=3.6 in]{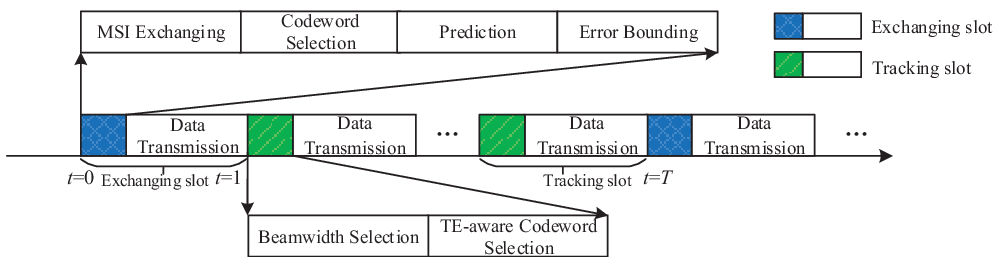}
\caption{Our frame structure design for high-mobility UAV mmWave networks. Therein, exchanging slot contains MSI exchanging, codeword selection, prediction, error bounding, and data transmission, while tracking slot contains beamwidth selection, TE-aware codeword selection, and data transmission.}
\label{figpro}
\end{figure}

Thanks to the integrated sensors, such as inertial measurement unit (IMU) and global position system (GPS), the UAV is able to derive its own MSI. \textcolor{black}{However, the r-UAV also needs the MSI of all t-UAVs and each t-UAV needs the r-UAV's MSI for \textcolor{kkb}{beam tracking}, which is challenging for the r-UAV/t-UAVs.}
%it is difficult for r-UAV/t-UAV to obtain the t-UAVs'/r-UAV's MSI which is also important for tracking the beam angles.
The GP-based MSI prediction is proposed to solve the problem in~\cite{25}.
\textcolor{black}{Specifically, the r-UAV/t-UAV's historical MSI is first exchanged with the t-UAV/r-UAV over a lower-frequency band and then the t-UAV will predict the future MSI of the r-UAV based on the historical MSI by using the GP-based MSI prediction model.}
However, the MSI prediction error causes the beam tracking error, which has a negative effect on the sum SE of UAV mmWave network and is not addressed by \cite{25}. In this paper, a new TE-aware transmission protocol is proposed to solve the problem as shown in Fig.~\ref{figpro}.
A conceptual frame structure is designed which contains two types of time slots. One is the exchanging slot (e-slot) and the other is the tracking slot (t-slot). Let us first focus on the e-slot. It is assumed that UAVs exchange MSI every $T$ t-slots, i.e., in an e-slot, to save resource for payload transmission. In the MSI exchanging period of the e-slot $t$, the r-UAV exchanges its historical MSI with each t-UAV and the t-UAV only exchanges its historical MSI with r-UAV over the low-rate control links that work in the lower-frequency band~\cite{lowfre}. Then t-UAVs and r-UAV perform codeword selection. Employing the GP-based MSI prediction algorithm proposed in~\cite{25}, each t-UAV predicts the MSI of r-UAV and r-UAV predicts the MSI of all t-UAVs in the next $T$ t-slots. In the tracking error bounding period, the UAVs estimate the TE of AOAs and AODs based on the GP prediction error. Compared to e-slot, t-slot does not have the MSI exchanging, prediction and error bounding, but has the TE-aware codeword selection. Specifically, in t-slot the t-UAVs and r-UAV achieve the adaptive beamwidth control against AODs/AOAs prediction errors by employing the TE-aware codeword selection. Compared to the motion-aware protocol in~\cite{25}, the new TE-aware protocol can be applied to the UAV mmWave network with higher mobility including random trajectories and high velocity. Since the new TE-aware protocol contains the error bounding and TE-aware codeword selection periods, it is able to deal with the beam tracking error caused by high mobility of UAVs. Next, we will detail how to bound the TE and how to select the proper codeword with suitable beamwidth against the TE in the following subsections.

\subsection{Tracking Error Bounding}
\label{secIV.B}
The tracking error of beam angles has a negative influence on the beam gain obtained by CCA. The proposed tracking error bounding algorithm uses the position/attitiude prediction error of the GP-based MSI prediction to obtain the beam angle tracking error, wherein the geometry relationship between UAVs and the Monte-Carlo method is utilized. \textcolor{black}{First, the algorithm used to predict MSI in the next $T$ t-slots after MSI exchanging is introduced. Due to the movement inertia, \textcolor{kkb}{the MSI between adjacent slots is correlated with each other}. Hence, the historical MSI can be used to predict the future MSI. According to the GP-based MSI prediction algorithm, the predicted position and attitude are estimated by the mean of the predictive distribution of the outputs (the future MSI) on the specific test dataset.} The predictive distribution of the output (the future MSI) is given by
\begin{eqnarray}
\begin{cases}\label{pdf}
 & ({{\boldsymbol{Y}}^{*}}|{{\boldsymbol{X}}^{*}},\mathcal{D})\sim\mathcal{N}(\boldsymbol{m}(\boldsymbol{Y}^{*}),\boldsymbol{K}(\boldsymbol{Y}^{*}))\\
\boldsymbol{m}(\boldsymbol{Y}^{*}) & =\boldsymbol{K}({{\boldsymbol{X}}^{*}},\boldsymbol{X})\boldsymbol{K}{{(\boldsymbol{X},\boldsymbol{X})}^{-1}}\boldsymbol{Y},\\
\boldsymbol{K}(\boldsymbol{Y}^{*}) & =-\boldsymbol{K}({{\boldsymbol{X}}^{*}},\boldsymbol{X})\boldsymbol{K}{{(\boldsymbol{X},\boldsymbol{X})}^{-1}}\boldsymbol{K}(\boldsymbol{X},{{\boldsymbol{X}}^{*}})\\
 & \thinspace\thinspace\thinspace\thinspace\thinspace+\boldsymbol{K}({{\boldsymbol{X}}^{*}},{{\boldsymbol{X}}^{*}})
\end{cases}
\end{eqnarray}
where $\boldsymbol{Y}^{*}$ is the output of the test dataset, $\boldsymbol{X}^{*}$ is the input of the test dataset, $\mathcal{D}=\{\boldsymbol{X},\boldsymbol{Y}\}$ is the training dataset, $\boldsymbol{m}(\boldsymbol{Y}^{*})$ is a mean matrix of the test data whose $i$-th element is the mean function $m(\boldsymbol{y}^{*}_i)$, and $\boldsymbol{K}(\boldsymbol{X},\boldsymbol{X})$ is a covariance matrix whose element is the kernel function $\boldsymbol{K}(i,j)=k(\boldsymbol{x}_i,\boldsymbol{x}_j)$. The input matrix of the training data is $\boldsymbol{X}=[{\boldsymbol{x}}_{1},\ldots,{\boldsymbol{x}}_{K}]^T$, where ${\boldsymbol{x}}_{k}$ is the input of the $k$-th training data. The output matrix of the training data is $\boldsymbol{Y}=[{\boldsymbol{y}}_{1},\ldots,{\boldsymbol{y}}_{K}]^T$, where ${\boldsymbol{y}}_{k}$ is the output of the $k$-th training data. The input and output of the test dataset $\boldsymbol{X}^{*}$ and $\boldsymbol{Y}^{*}$ can be defined similarly.

Next, the specific training dataset $(\boldsymbol{X},\boldsymbol{Y})$ and the test dataset $(\boldsymbol{X}^{*},\boldsymbol{Y}^{*})$ for MSI prediction and tracking error bounding are introduced. We focus on the error bounding for the x-coordinate of UAV position and the yaw angle of UAV attitude to explain the general idea, and other positions and attitudes can be developed similarly. Let us first define the inputs and outputs of the $k$-th training data for the x-coordinate of r-UAV prediction as column vectors $\boldsymbol{x}_k=\{x_r(t-T),\ldots,x_r(t-1)\}$ and $\boldsymbol{y}_k=\{x_{r}(t),\ldots,x_{r}(t+T_f)\}$, respectively, where $T_f$ represents the number of slots predicted ahead of the reference time $t$. Define the inputs and outputs of the $k$-th training data for UAV yaw angle prediction as $\boldsymbol{x}_k=\{\phi_r(t-T),\ldots,\phi_r(t-1),\boldsymbol{v}_{r}(t),\boldsymbol{a}_{r}(t)\}$ and
$\boldsymbol{y}_k=\{\phi_r(t),\ldots,\phi_r(t+T_f)\}$, and the inputs and outputs of the $q$-th test data can be defined in a similar way. Then we continue to bound the tracking error range. Note that the covariance represents the statistical uncertainty of the prediction. According to the properties of Gaussian distribution, the $99\%$ confidence region can be bounded by adding and subtracting three times of the standard deviation to/from the mean value. That is to say, the GP-based MSI prediction algorithm can bound the prediction error within a certain range of a confidence value. More specifically, the covariance of the output of the $q$-th test data is written by $\boldsymbol{k}(\boldsymbol{y}^{*}_{p})=\{k({y^{*}(t)}),\ldots,k({y^{*}(t+T_f)})\}$. Then the predicted x-coordinate of r-UAV in the slot $t$ is $\hat{x}_{r}(t)=m({x}_{r}(t))$. The real position on the x-coordinate obeys the Gaussian distribution with the mean $m({x}_{r}(t))$ and the covariance $k(x_r(t))$. It belongs to the range $[m({x}_{r}(t))-3\sqrt{k(x_{r}(t))},m({x}_{r}(t))+3\sqrt{k(x_{r}(t))}]$ which is called the error range of x-coordinate with $99\%$ confidence. The distribution of the yaw angle can be given in a similar way.

The predictive MSI $(\boldsymbol{X}_r(t),\boldsymbol{\Theta}_r(t))$ and the corresponding error range is obtained after GP-based prediction. Meanwhile, the UAV's real position is assumed to follow the Gaussian distribution with the mean $\boldsymbol{m}(\boldsymbol{X}_r(t))=[m(x_r(t)),m(y_r(t)),m(z_r(t))]$ and covariance $\boldsymbol{K}(\boldsymbol{X}_r(t))=[k(x_r(t)),k(y_r(t)),k(z_r(t))]$ for the prediction problem, the real UAV's attitude is also assumed to obey the Gaussian distribution with the mean $\boldsymbol{m}(\boldsymbol{\Theta}_r(t))=[m(\psi_r(t)),m(\theta_r(t)),m(\phi_r(t))]$ and covariance $\boldsymbol{K}(\boldsymbol{\Theta}_r(t))=[k(\psi_r(t)),k(\theta_r(t)),k(\phi_r(t))]$.
The Monte-Carlo method is used to estimate the beam angle error. To achieve this goal, each UAV generates two random variables $\overline{\boldsymbol{X}}_{r}(t)\sim\mathcal{N}(\boldsymbol{m}(\boldsymbol{X}_r(t)),\boldsymbol{K}(\boldsymbol{X}_r(t)))$ and $\overline{\boldsymbol{\Theta}}_{r}(t)\sim\mathcal{N}(\boldsymbol{m}(\boldsymbol{\Theta}_r(t)),\boldsymbol{K}(\boldsymbol{\Theta}_r(t)))$ with reference to the GP-predicted distributions.
The corresponding beam angles $\overline{\alpha}(t),\overline{\beta}(t)$ are calculated by the geometry based beam tracking algorithm proposed in~\cite{25}. \textcolor{black}{Let us focus on the transmitting beam angle of a t-UAV to explain the \textcolor{kkb}{basic idea, and} other beam angles can be calculated similarly. First, we establish different coordinate frames, including the global coordinate frame (g-frame), the t-UAV coordinate frame (a-frame), the r-UAV coordinate frame (b-frame), and the antenna array coordinate frame (c-frame). Then, the coordinate frame transformation can be obtained by the predicted MSI of the r-UAV and the MSI of t-UAVs given by sensors, such as IMU, GPS, and so on. Finally, the transmitting beam angle of the t-UAV can be calculated in c-frame by using the geometric relationship between the CCA of the r-UAV and the CCA of t-UAVs.}
\textcolor{black}{Repeat the above process and a lot of} \textcolor{black}{samples of beam angles will be generated.
Then the empirical probability
distribution functions (EDFs) of beam angles and the mean value $\hat{\alpha}_{\text{mean}}(t)$ and $\hat{\beta}_{\text{mean}}(t)$ are obtained based on the statistics over the set of samples $\overline{\alpha}(t)$ and $\overline{\beta}(t)$. Subsequently the error range of the azimuth angle $[\hat{\alpha}_{\text{min}},\hat{\alpha}_{\text{max}}]$ can be derived with reference to the following probability
\begin{eqnarray}
\label{palpha}
\text{Pr}(\overline{\alpha}(t)\in[\hat{\alpha}_{\text{min}}(t),\hat{\alpha}_{\text{max}}(t)])=\text{P}_{\alpha},
\end{eqnarray}
and $[\hat{\alpha}_{\text{min}},\hat{\alpha}_{\text{max}}]$ is uniquely determined by assuming the range is centered by the mean value of $\overline{\alpha}(t)$, i.e., $(\hat{\alpha}_{\text{min}}+\hat{\alpha}_{\text{max}})/2=\hat{\alpha}_{\text{mean}}$.} In a similar way, the error range of the elevation angle $[\hat{\beta}_{\text{min}},\hat{\beta}_{\text{max}}]$ can be derived with reference to the following probability
\begin{eqnarray}
\label{pbeta}
\text{Pr}(\overline{\beta}(t)\in[\hat{\beta}_{\text{min}}(t),\hat{\beta}_{\text{max}}(t)])=\text{P}_{\beta}.
\end{eqnarray}
In summary, the error bounding algorithm is shown in Algorithm~\ref{AL2}, and the corresponding complexity analysis is given below.

\emph{Complexity Analysis}: For each slot, the complexity of the prediction with trained GP model is $\mathcal{O}(N_K)$, where $N_K$ is the dimension of kernel matrix $N_K$. Generating $\overline{\boldsymbol{X}}_{r}(t)$, $\overline{\boldsymbol{\Theta}}_{r}$ and computing the corresponding beam angles takes $\mathcal{O}(I_{\text{max}})$ operations. Computing the EDF of the beam angles and obtaining their error ranges takes $\mathcal{O}(I_{\text{max}})$ operations. Consequently, the complexity of Algorithm~\ref{AL2} at each slot is $\mathcal{O}(N_K+I_{\text{max}})$.

\begin{algorithm}[t]
\caption{UAV Position-Attitude Prediction Error Bounding Algorithm}
\label{AL2}
\begin{algorithmic}[1]
\STATE Optimize the hyperparameters in the GP model using the training data.
\FOR {$t_0=1:T:T_{\text{max}}$}
\STATE Given MSI before $t$, predict $X_r(t)$, $X_t(t)$, ${\boldsymbol{\Theta }}_{r}(t)$ and ${\boldsymbol{\Theta }}_{t}(t)$, derive the covariation $\boldsymbol{K}(\boldsymbol{X}_r(t))$ and $\boldsymbol{K}(\boldsymbol{\Theta}_r(t))$, $t\in [t_0,t_0+T]$.
\FOR {each $t\in [t_0,t_0+T]$}
\FOR {$i=1:I_{\text{max}}$}
\STATE Generate $\overline{\boldsymbol{X}}_{r}(t)\sim\mathcal{N}(\boldsymbol{m}(\boldsymbol{X}_r(t)),\boldsymbol{K}(\boldsymbol{X}_r(t)))$, $\overline{\boldsymbol{\Theta}}_{r}(t)\sim\mathcal{N}(\boldsymbol{m}(\boldsymbol{\Theta}_r(t)),\boldsymbol{K}(\boldsymbol{\Theta}_r(t)))$.
\STATE Compute $\overline{\alpha}_i(t),\overline{\beta}_i(t)$ using the geometry based beam tracking algorithm in~\cite{25}.
\ENDFOR
\STATE Compute the PDF of the beam angles based on $\overline{\alpha}_i(t),\overline{\beta}_i(t), 1\leq i\leq I_{\max}$.
\STATE $[\hat{\alpha}_{\text{min}},\hat{\alpha}_{\text{max}}]$ and $[\hat{\beta}_{\text{min}},\hat{\beta}_{\text{max}}]$ according to (\ref{palpha}) and (\ref{pbeta}).
\ENDFOR
\ENDFOR
\end{algorithmic}
\end{algorithm}

\subsection{Tracking-Error-Aware Codeword Selection with 3D Beamwidth Control}
\label{secIV.C}
At the r-UAV side, in the presence of beam tracking error for a t-UAV,
the optimal beamwidth is not always the narrowest one that makes full
use of all the possible DREs. \textcolor{black}{A wider beamwidth (a smaller subarray size) should be selected to improve the beam gain with \textcolor{kkb}{the increase of TE}. \textcolor{kkb}{However, too wide beamwith (too small subarray size) can cause the beam gain reduction} due to the decrease in beam directivity. Therefore, the appropriate beamwidth (subarray size/layer in the codebook) should be selected.}
%Therefore, an appropriate beamwidth needs to be selected to optimize the beam gain against the tracking error.
To this end, the TE-aware codeword selection using the multi-resolution CCA codebook is proposed to
achieve adaptive 3D beamwidth control for more robust beam tracking
at the r-UAV side.

Without loss of generality, let us focus on the TE-aware codeword
selection for the $k$-th t-UAV at the r-UAV side. \textcolor{black}{The beam gain is \textcolor{kkb}{selected} as the optimization objective, and the problem of beamwidth control is translated to choose the appropriate subarray size, which \textcolor{kkb}{corresponds to} the appropriate layer in the codebook, to optimize the beam gain. A two-step scheme is proposed to find the suboptimal layer index in the codebook and the codewords can be selected at the same time.} Specifically, to start with,
r-UAV can get the estimation of $(\hat{\alpha}_{k},\hat{\beta}_{k})$ according to~(\ref{pdf})
and obtain the subarray-dependent $MN\times1$ combining vector $\ensuremath{\boldsymbol{w}}_{k}(\hat{\alpha}_{k},\hat{\beta}_{k},\mathcal{S}_{k}^{\mathrm{r}})$.
Then the corresponding beam gain along an arbitrary angle $(\alpha,\beta)$
is expressed as
\begin{eqnarray}
G_{k}(\alpha,\beta,\ensuremath{\boldsymbol{w}}_{k})=\sqrt{N_{\text{act}}M_{\text{act}}}\boldsymbol{a}_{k}(\alpha,\beta)^{H}\ensuremath{\boldsymbol{w}}_{k}\left(\hat{\alpha}_{k},\hat{\beta}_{k},\mathcal{S}_{k}^{\mathrm{r}}\right),\label{beamgain}
\end{eqnarray}
where $N_{\text{act,}k}M_{\text{act},k}$ is the size of the activated
subarray $\mathcal{S}_{k}^{\mathrm{r}}$ to support $\boldsymbol{v}$,
and $\boldsymbol{a}_{k}(\alpha,\beta)$ is the array steering vector
along the angle $(\alpha,\beta)$ (cf.(1)). When the estimated elevation
angle $\hat{\beta}_{k}$ is fixed, the minimum beam gain achieved by CCA on
the azimuth plane with the azimuth angle error range $[\alpha_{k,\mathrm{min}},\alpha_{k,\mathrm{max}}]$
is defined as $G_{k}(\ensuremath{\boldsymbol{w}}_{k})_{\text{a,min}}=\min\{G_{k}(\alpha_{k,\mathrm{min}},\hat{\beta}_{k},\ensuremath{\boldsymbol{w}}_{k}),G_{k}(\alpha_{k,\mathrm{max}},\hat{\beta}_{k},\ensuremath{\boldsymbol{w}}_{k})\}$.
Similarly, given the estimated azimuth angle $\hat{\alpha}_{k}$,
the minimum beam gain on the elevation plane with the elevation angle
error range $[\beta_{k,\mathrm{min}},\beta_{k,\mathrm{max}}]$ is
defined as $G_{k}(\ensuremath{\boldsymbol{w}}_{k})_{\text{e,min}}=\min\{G_{k}(\hat{\alpha}_{k},\beta_{k,\mathrm{min}},\ensuremath{\boldsymbol{w}}_{k}),G_{k}(\hat{\alpha}_{k},\beta_{k,\mathrm{max}},\ensuremath{\boldsymbol{w}}_{k})\}$.
Since the beamwidth is determined by the size of the activated subarray
$\mathcal{S}_{k}^{\mathrm{r}}\left(m_{s,k},n_{s,k},\ensuremath{\boldsymbol{p}}_{k}\left(\hat{\alpha}_{k}\right)\right)$,
the beamwidth control is translated to the codeword selection with
a proper subarray size $(M_{\text{act},k}=m_{s,k},\thinspace N_{\text{act,}k}=n_{s,k})$.

\textcolor{black}{Note that the beamwidth on the azimuth plane and elevation plane is
 mainly determined by $n_{s,k}$ and $m_{s,k}$ jointly. For feasibility, we propose a suboptimal two-step scheme to find the codeword layer indexes $n^*_{s,k}$ and $m^*_{s,k}$, sequentially.} More specifically, let us define $\mathcal{M}(\mathcal{V}_{r})$ and $\mathcal{N}(\mathcal{V}_{r})$
as the sets of all layers' indexes $\left\{ \ensuremath{m_{s}}\right\}$
and $\left\{ n_{s}\right\} $ in the CCA codebook used by the r-UAV. Given $M_{\text{max},k}:=\text{max}\thinspace\left\{ \mathcal{M}(\mathcal{V}_{r})\right\} $
and $\forall n_{s,k}\in\mathcal{N}(\mathcal{V}_{r})$, the optimal
codeword $\boldsymbol{v}(i_{k}^{*},j_{k}^{*},\mathcal{S}\left(M_{\text{max},k},n_{s,k}\right)):=\boldsymbol{v}^{*}\left(n_{s,k}|M_{\text{max},k}\right)$
can be selected based on the predicted beam angles $(\hat{\alpha_{k}},\hat{\beta_{k}})$
according to (\ref{solutionr}). Then $G_{k}\left(\boldsymbol{v}^{*}\left(n_{s,k}|M_{\text{max},k}\right)\right)_{\text{a,min}}$
can be calculated for all possible $n_{s,k}\in\mathcal{N}(\mathcal{V}_{r})$,
and the one with the largest beam gain can return the optimal $n_{s,k}^{*}$.
Now, with the fixed $n_{s,k}^{*}$ and $\forall m_{s,k}\in\mathcal{M}(\mathcal{V}_{r})$, we can get $\boldsymbol{v}(i_{k}^{*},j_{k}^{*},\mathcal{S}(m_{s,k},n_{s,k}^{*})):=\boldsymbol{v}^{*}(m_{s,k}|n_{s,k}^{*})$,
and by finding the largest $G_{k}(\boldsymbol{v}^{*}(m_{s,k}|n_{s,k}^{*}))_{\mathrm{e,min}}$
in $\mathcal{M}(\mathcal{V}_{r})$, we can get the optimal $m_{s,k}^{*}$ given $n_{s,k}^{*}$.
%\textcolor{black}{The suboptimal solution $(m_{s,k}^{*},n_{s,k}^{*})$ has been found.}
In summary, the TE-aware CCA codeword selection algorithm is given by Algorithm~\ref{AL3}, and the corresponding complexity analysis is given as follows.%\textcolor{black}{\emph{Remark2}:
\textcolor{black}{
\begin{remark}
\label{Remark2}
%\textcolor{black}{A wider beamwidth (a smaller subarray size) should be selected to improve the beam gain with increasing TE. However, too wider beamwith (too smaller subarray size) can also cause the beam gain reduction due to the decrease in beam directivity. Therefore, the appropriate beamwidth (subarray size/layer in the codebook) should be selected.}
\textcolor{black}{The maximum-resolution layer of the multi-resolution CCA codebook with the fixed minimum beamwidth is used in Algorithm~\ref{AL1} for the optimal solution.}
%The search space of Algorithm~\ref{AL1} for optimal solution is the maximum-resolution layer of the multi-resolution CCA codebook with the fixed minimum beamwidth.
To solve the TE-aware beam tracking problem, the search space of Algorithm~\ref{AL3} is the whole multi-resolution CCA codebook since the beamwidth is selected according to the estimated TE.
\end{remark}}
\textcolor{black}{\emph{Complexity Analysis}: The complexity of searching for optimal $n_s$ and $m_s$ in our codebook is $\mathcal{O}((|\mathcal{M}(\mathcal{V})|+|\mathcal{N}(\mathcal{V})|)M_{\text{act}}N_{\text{act}}K)$, \textcolor{kkb}{where $|\cdot|$ is the cardinality of a set;} the complexity of the updating procedure is similar with Algorithm~1. \textcolor{kkb}{In total,} the complexity of Algorithm~\ref{AL3} at each slot is $\mathcal{O}((|\mathcal{M}(\mathcal{V})|+|\mathcal{N}(\mathcal{V})|)M_{\text{act}}N_{\text{act}}K+K^3)$ in the worst case.}

\begin{algorithm}[h]
\caption{Tracking-Error-Aware codeword selection with 3D Beamwidth Control}
\label{AL3} \begin{algorithmic}[1]
\INPUT $(\hat{\alpha_{k}},\hat{\beta}_{k})$,
$[\hat{\alpha}_{\text{min},k},\hat{\alpha}_{\text{max},k}]$, $[\hat{\beta}_{\text{min},k},\hat{\beta}_{\text{max},k}]$,
$\mathcal{V}_{r}$ \INIT $M_{\text{max},k}=\text{argmax}_{m\in\mathcal{M}(\mathcal{V}_{r})}\thinspace m$.
\FOR {$n_{s,k}\in\mathcal{N}(\mathcal{V}_{r})$}
\STATE Select the optimal codeword $\boldsymbol{v}^{*}(n_{s,k}|M_{\text{max},k})$ in codebook $\mathcal{V}_{k}$ according to (\ref{solutionr}).
\STATE Calculate $G_{k}(\boldsymbol{v}^{*}(n_{s,k}|M_{\text{max},k}))_{\text{a,min}}$.
\ENDFOR
\STATE $n_{s,k}^{*}=\text{arg max}_{n_{s,k}\in\mathcal{N}(\mathcal{V}_{r})}G(\boldsymbol{v}^{*}(n_{s,k}|M_{\text{max},k}))_{\text{a,min}}$.
\FOR {$m_{s,k}\in\mathcal{M}(\mathcal{V}_{r})$}
\STATE Given
$n_{s,k}^{*}$, select the optimal codeword $\boldsymbol{v}^{*}(m_{s,k}|n_{s,k}^{*})$
in $\mathcal{V}_{k}$ according to (\ref{solutionr}).
\STATE Calculate
$G_{k}(\boldsymbol{v}^{*}(m_{s,k}|n_{s,k}^{*}))_{\mathrm{e,min}}$.
\ENDFOR
\STATE $m_{s,k}^{*}=\text{arg max}_{m_{s,k}\in\mathcal{M}(\mathcal{V}_{r})}G_{k}(\boldsymbol{v}^{*}(m_{s,k}|n_{s,k}^{*}))_{\mathrm{e,min}}$.
\STATE Codeword $\boldsymbol{w}_{k}(\mathcal{S}_{k}^{r})=\boldsymbol{v}(i_{k}^{*},j_{k}^{*},\mathcal{S}(m_{s,k}^{*},n_{s,k}^{*})$
is selected. \STATE Detect confliction and update $\boldsymbol{w}_{k}(\mathcal{S}_{k}^{r})$
by using Algorithm~\ref{AL1}.
\OUTPUT $\boldsymbol{w}_{k}(\mathcal{S}_{k}^{r})$.
\end{algorithmic}
\end{algorithm}

%ȱһ���������㷨������Beamselection ��SPCS�Ĺ�ϵ
\section{Simulation}
\label{Sec5}
 In this section, numerical results are provided to evaluate the effectiveness of the proposed codebook based SPAS algorithm and TE-aware beamwidth control for beam tracking. The simulation setups are given as follows. The UAV mmWave network with a carrier frequency 60 GHz is considered and the carrier wavelength is ${\lambda }_{c}=0.005$~m. The inter-UAV mmWave channel follows the model in (\ref{channel}). The Smooth-Turn mobility model~\cite{Smooth-Turn} is used to generate the UAV's trajectory on the $xy$-plane, where the mean of the duration is set as $1/{\lambda}=1$~s and the variance is set as $\sigma_r^2=0.05$.
 %The distance between the t-UAV and r-UAV is limited to no more than $D_{\text{r,max}}=40$~m and
 The distance between UAVs are limited to no less than $D_{\text{r,min}}=10$~m. Referring to the composite wing UAV of CHC P316 technical parameters~\cite{HUAC}, the horizontal velocity is no more than $v_{xy}=20$~m/s, the minimum and maximum vertical velocities are set as $v_{t(r),z,\text{min}}=2$~m/s and $v_{t(r),z,\text{max}}=3$~m/s, respectively. The time slot duration is set as $\delta{t}=10$~ms. Thus, the horizontal distance of the UAV navigation is approximately no more than $0.2$~m which can be almost neglected during the prediction process.
The UAV MSI exchanging period is set as $T=50$, i.e., 500~ms.

 \textcolor{black}{%In practice, the CCA radius $R_{\text{cyl}}$ determined by UAV manufacturing and the corresponding the size of CCA may be quite large to conform the surface of the UAV.
 In the simulation, the size of the t-UAV's DRE-covered CCA is set as $N_t=64$ and $M_t=16$ for the consideration of the computational complexity. Meanwhile, the radius $R_{\text{cyl}}$ is set as $R_{\text{cyl}}=0.0509$~m to achieve the desired beam pattern and the array response.} Hence, the maximum number of the activated elements on the $xy$-plane is $N_{\text{act,max}}=21$. The radiation range of the directive elements on the azimuth plane and the elevation plane is set as $\Delta\alpha=\frac{2\pi}{3}$ and $\Delta\beta=\pi$. %In this case, (\ref{cvlayera}) and (\ref{cvlayere}) hold and the CCA can provide the omni-directional beam.
The specific radiation range of the DRE on the azimuth plane and the elevation plane $(m,n)$ for CCA is written as
\begin{eqnarray}
[\alpha_{n,\text{min}},\alpha_{n,\text{max}}]\!=\!\left[\text{mod}(\phi_{n}\!-\!\frac{\pi}{3},2\pi),\text{mod}(\phi_{n}\!+\!\frac{\pi}{3},2\pi)\right],
\end{eqnarray}
where $\phi_{n}=\frac{2n-1-N}{2}\Delta\phi$ and $[\beta_{m,\text{min}},\beta_{m,\text{max}}]=[0,\pi]$. For comparison, the radiation range of the DRE for the UPA is $[\alpha_{n,\text{min}},\alpha_{n,\text{max}}]=[-\frac{\pi}{3},\frac{\pi}{3}]$ and $[\beta_{m,\text{min}},\beta_{m,\text{max}}]=[0,\pi]$.
\begin{figure}[!t]
\centering
\includegraphics[scale=0.8, width=2.6in]{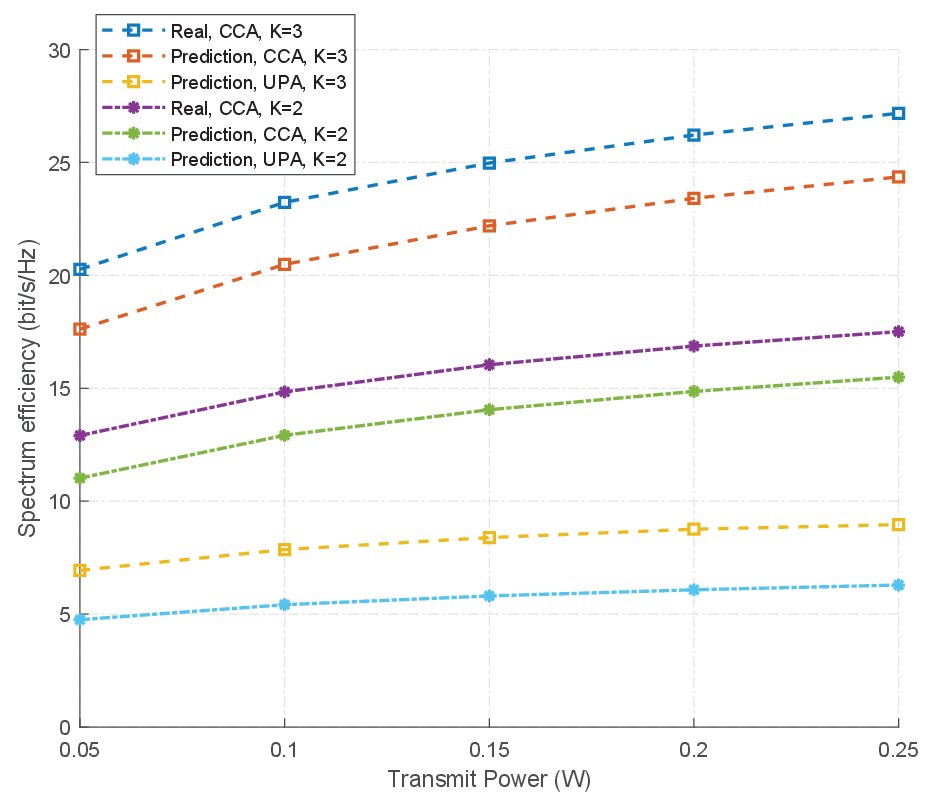}%Up
\caption{\textcolor{black}{Spectral efficiency vs. transmit power with different numbers of users.}}
\label{figsimulation1}
\end{figure}
% Simulation parameters is shown in Table I.

\subsection{Codebook Based Beam Tracking: CCA Versus UPA}
\begin{figure}[!t]
\centering
\includegraphics[scale=0.8, width=2.6in]{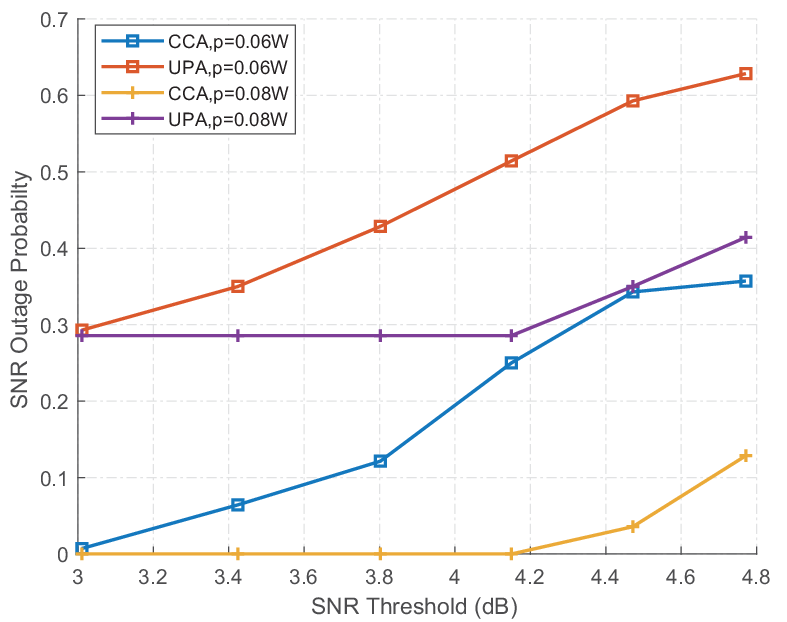}
%\caption{The SINR outage vs. SINR threshold with $K=3$ t-UAVs and the transmit power $p_k=0.01$~W, $N_{\text{act,max}}=10$.}
\caption{The SNR outage vs. SNR threshold with $K=2$ t-UAVs, the transmit power $p_k=0.06$~W and $p_k=0.08$~W, $N_{\text{act,max}}=10$.}
\label{figsimulation2}
\end{figure}
In this subsection, two beam tracking schemes with different types of antenna array are illustrated by simulation results. One is the proposed DRE-covered CCA scheme where all the t-UAVs are equipped with the CCA of the size $N_t=64$, $M_t=16$, and the r-UAV is equipped with the CCA of the size $N_r=64$, $M_r=112$. The other is the DRE-covered UPA scheme where all the t-UAVs are equipped with the UPA of the size $N_t=64$, $M_t=16$ and the r-UAV is equipped with UPA of the size $N_r=64$, $M_r=112$. The AOAs and AODs are predicted by using the GP-based position-attitude prediction algorithm in the prediction schemes. The real AOAs and AODs are used in the real schemes. Without the TE-aware beamwidth control, the initial size of subarray is selected as $M=M_{\text{act,max}}$ and $N=N_{\text{act,max}}$. In the CCA scheme, the codeword selection algorithm is performed by t-UAVs and r-UAV. In the UPA scheme, the UPA is equally partitioned into subarrays with the size of $M_r/K \times N_r$ and the beamforming/combining vector is calculated by the UAV position-attitude prediction based beam tracking algorithm in~\cite{25}. The SE is calculated according to (\ref{SE2}).

Fig.~\ref{figsimulation1} shows the sum SE against transmitting power with different number of t-UAVs. It is observed that the codeword selection algorithm is effective
and the CCA scheme achieves higher SE than the UPA scheme obviously with different t-UAV number $K$. The main reason is that the UPA with DREs can only receive/transmit the signal within a limited angular range at a certain time slot while the CCA does not have such limitation. It is also shown that the gap between the real schemes and the prediction schemes is small. Thus, the codebook-based beam tracking algorithm is effective in the considered CCA-enabled UAV mmWave network.

Next, we compare the outage probabilities achieved by the DER-covered CCA and UPA schemes in the UAV mmWave network. Here, the network outage probability is defined as~\cite{6}
\begin{eqnarray}
\label{out}
\text{Pr}(\underset{1\leq k \leq K}\min{\text{SNR}_k(t)}<\text{SNR}_{\text{th}}),
\end{eqnarray}
where $\text{SNR}_{\text{th}}$ is a certain SNR threshold. If an arbitrary link between the t-UAV and the r-UAV is in the outage, there is an outage in the network. Hence, the outage probability is determined by the minimum SNR among $K$ t-UAVs. Fig.~\ref{figsimulation2} shows the outage probability against the SNR threshold to further demonstrate the coverage of the CCA scheme and the UPA scheme with $K=2$, and the fixed power $p_k=0.06$~W and $p_k=0.08$~W. It is shown that the CCA scheme is always superior to the UPA scheme in the coverage. In some cases, the outage remains unchanged with the threshold increasing. The reason is that the outage in these cases is mainly determined by the coverage ability of the arrays.
%Fig.~\ref{figsimulation2} shows the outage probabilities against the SINR threshold to further demonstrate the coverage advantages of the CCA over the the UPA, where the fixed power $p_k=0.01$~W and t-UAV number $K=3$ are set. It is shown that the CCA scheme is superior to the UPA scheme in terms of coverage capability.
\begin{figure}[!t]
\centering
\includegraphics[scale=0.8, width=2.5in]{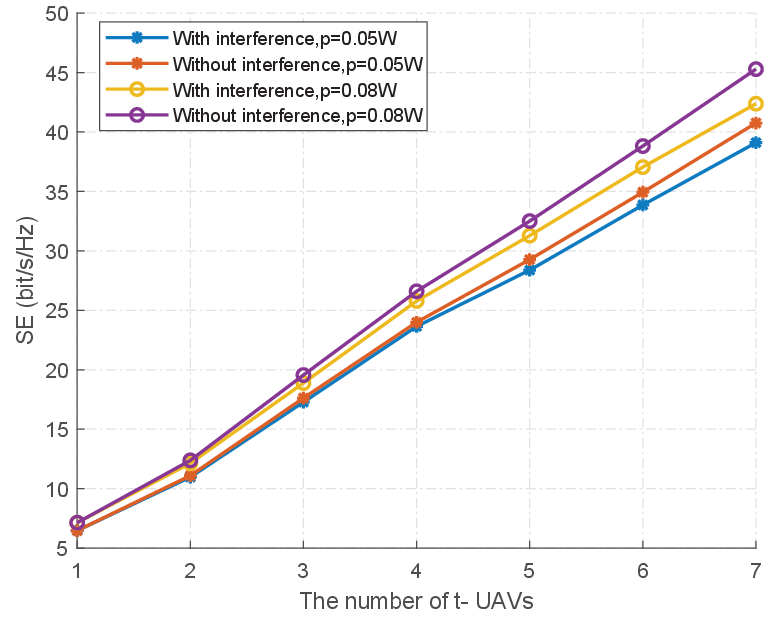}
\caption{\textcolor{black}{The spectral efficiency vs. the number of t-UAVs with different transmit power.}}
\label{figsimulation3}
\end{figure}

\begin{figure}[!t]
\centering
\includegraphics[scale=0.8, width=2.5in]{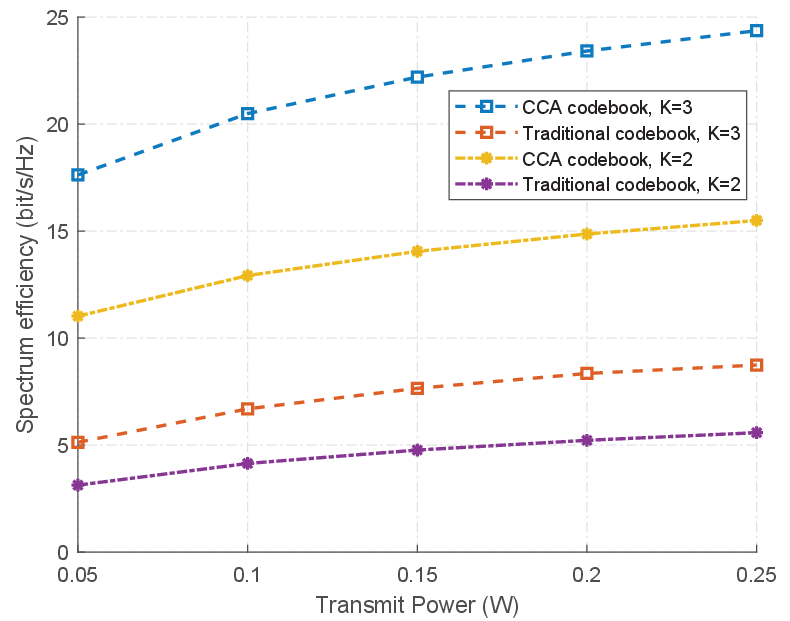}
\caption{\textcolor{black}{The spectral efficiency vs. transmit power with different codebooks.}}
\label{figsimulationcode}
\end{figure}
\textcolor{black}{In this paper, we mainly focus on the analog beam tracking without considering the inter-UAV interference. The sum SE calculated by (\ref{SE}) and (\ref{SE2}) with different numbers of t-UAVs and the given transmit power is shown in Fig.~\ref{figsimulation3}, respectively, to verify the influence of the inter-UAV interference. It is shown that the sum SE of the scheme without interference calculated by (\ref{SE2}) is similar with that of the scheme with interference calculated by (\ref{SE}) with the appropriate number of t-UAVs and the limited transmit power. The gap between the schemes increases as the power and the number of t-UAVs increase. Therefore, the inter-UAV interference can be neglected in the considered scenario.}

\textcolor{black}{As shown in Fig.~\ref{figsimulationcode}, the SE of the CCA codebook scheme and the traditional codebook scheme is compared. The proposed DRE-covered CCA codebook is used in the CCA codebook scheme. In the traditional codebook scheme, the codebook without subarray partition is used. The CCA on the r-UAV is equally partitioned into $K$ fixed subarrays with the size of $M_r/K\times N_r$ and the CCA on the t-UAVs is not partitioned. The CCA scheme performs better than the traditional scheme with different numbers of t-UAVs and different transmit power. The reason is that only a part of activated DREs' radiation range covers the determined radiation direction.
}
%In this paper, we mainly focus on the beam tracking with the analog beamforming. In fact, the digital baseband processing can be integrated easily to further suppress the multiuser interference as well. The additional channel state information needed by the digital baseband processing can be obtained via pilot-aided estimation within the links established by the analog beam. The sum SE of the CCA scheme with an MMSE baseband digital combiner, namely the Hybrid beamforming (BF) scheme, and the only analog beamforming scheme are illustrated in Fig.~\ref{figsimulation3}. When the number of t-UAVs $K$ increases, the sum SE of the analog BF scheme raises at first, and then decreases. Since the multiuser interference increases with the number of t-UAVs and the transmit power raising, the sum SE in the case $K=4$ is lower than the cases $K=3$ and $K=2$ when the transmit power is relatively high. It is observed that the SE of the hybrid BF scheme is higher than the analog BF scheme. The sum SE of hybrid BF scheme in the case of $K=4$ t-UAVs is high than the case of $K=3$ thanks to the MMSE-based multiuser interference mitigation.
%\begin{figure}[!t]
%\centering
%\includegraphics[scale=0.8, width=2.5in]{figure/hybrid.eps}
%\caption{The spectral efficiency vs. transmit power with different numbers of users.}
%\label{figsimulation3}
%\end{figure}

\subsection{TE-aware CCA Codeword Selection with 3D Beamwidth Control}
%\begin{figure}[!t]
%\centering
%\includegraphics[scale=0.8, width=2.5in]{figure/erangey.eps}
%\caption{The position prediction and the corresponding error range.}
%\label{figsimulation4}
%\end{figure}
\begin{figure}
\begin{minipage}[t]{0.5\linewidth}
\centering
\includegraphics[width=1.8in]{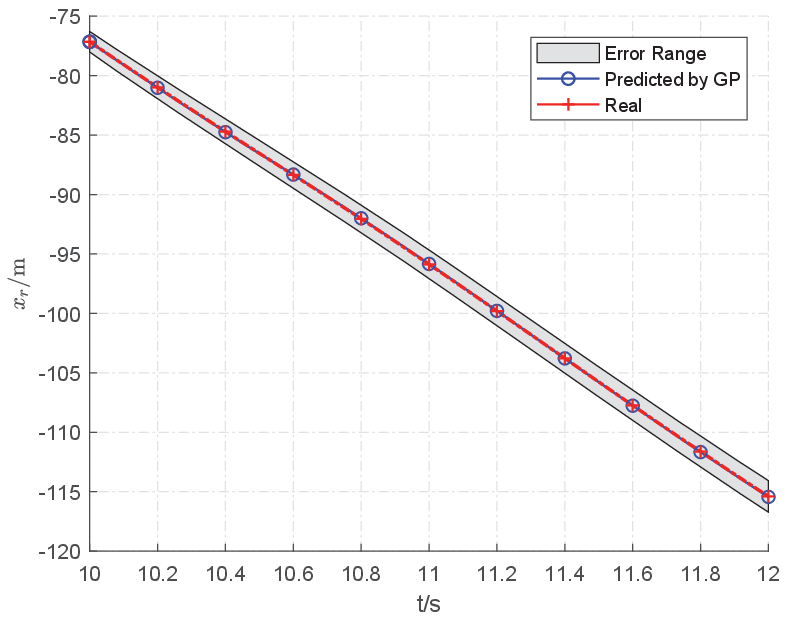}
\label{fig:side:a}
\end{minipage}%
\begin{minipage}[t]{0.5\linewidth}
\centering
\includegraphics[width=1.8in]{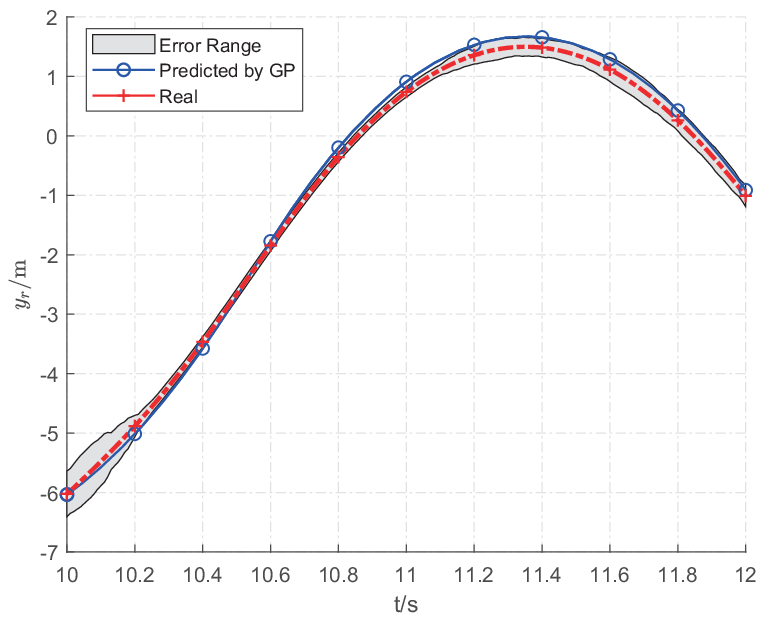}
\label{fig:side:b}
\end{minipage}
\caption{\textcolor{black}{The position prediction and the corresponding error range.}}
\label{figsimulation4}
\end{figure}
Although the GP-based UAV's position and attitude prediction results fit well with the position and attitude data, the prediction performance is effected by UAV's mobility. When the UAV has higher mobility such as the more random trajectory and high velocity, the prediction error may influence the beam tracking. The covariance of the turning radius is set as $\sigma_r^2=0.06$ which determines the randomness of the trajectory, and the velocity is set as no more than $v_{xy}=20$~m/s. The UAV's position and attitude prediction results and the corresponding error range is shown in Fig.~\ref{figsimulation4}, where the error range with $99\%$ confidence covers the error between the real data and the predicted data in most cases.

\begin{figure}[!t]
\centering
\includegraphics[scale=0.8, width=2.5in]{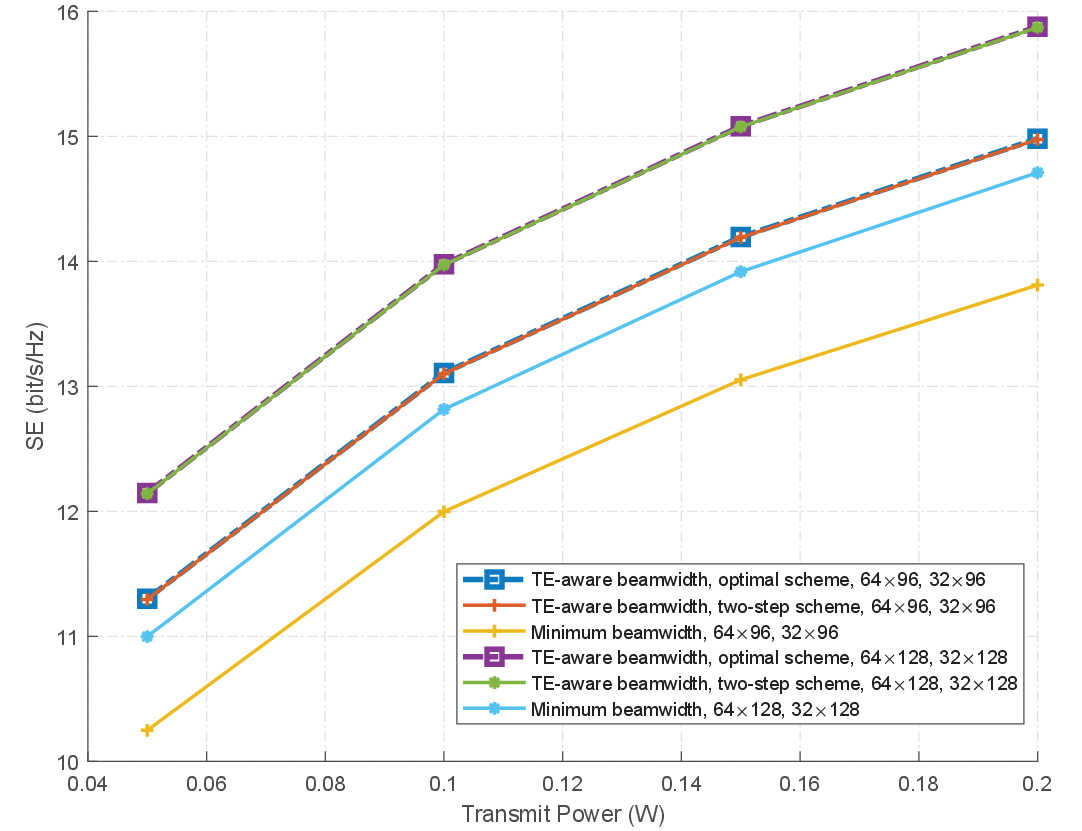}
%\caption{\textcolor{black}{Spectral efficiency vs. transmit power \textcolor{black}{under} TE-aware beam selection scheme and the minimum beamwidth scheme.}}
\caption{\textcolor{black}{Spectral efficiency vs. transmit power \textcolor{black}{under} TE-aware beam selection scheme.}}
%with the array sizes of $M_t\times N_t=32 \times 96$, $M_r \times N_r =64 \times 96$ and $M_t\times N_t=32 \times 128$, $M_r \times N_r =64 \times 128$.}}
\label{figsimulation5}
\end{figure}

The SEs of two array schemes against the transmit power with $K=2$ t-UAVs are illustrated in Fig.~\ref{figsimulation5}. The TE-aware codeword selection uses the proposed Algorithm~\ref{AL2} and Algorithm~\ref{AL3}. Serving as a reference, the minimum-beamwidth scheme always select the minimum beamwidth, i.e., the maximum number of antenna elements for an activated subarray. \textcolor{black}{\textcolor{kkb}{To evaluate the performance of the proposed two-step scheme, the exhaustive searching scheme for the optimal layer index is also simulated as a comparison}, where the traversal of all codebook layers is executed. As shown in Fig.~\ref{figsimulation5}, the sum SE of the TE-aware codeword selection scheme is better than the minimum-beamwidth codeword selection scheme. In addition, the curve of the two-step scheme almost overlaps that of the optimal scheme, as the tracking error of elevation angle is relatively small and the optimal $m$ varies in a relatively small range.}

\textcolor{black}{In order to verify the feasibility of the proposed scheme, the average latency in an e-slot/t-slot/frame duration with different UAV numbers are evaluated as follows:}

\textcolor{black}{
The total latency in e-slot is given by $t_{\text {total,e}}=t_{\text{MSI}}+t_{\text{tra}}+t_{\text{pro}}+t_{\text{local,e}}$ and the total latency in t-slot is $t_{\text {total,t}}=t_{\text{tra}}+t_{\text{pro}}+t_{\text{local,t}}$, where the transmission time of MSI $t_{\text{MSI}}=\frac{B_{\text{MSI}}}{C_{\text{LB}}}$, the transmission time with mmWave band $t_{\text{tra}}=\frac{B_{\text{data}}}{C_{\text{ave}}}$, the propagation time $t_{\text{pro}}=\frac{D_{k,\max}}{c}$\textcolor{kkb}{, and the local processing times $t_{\text{local,t}}$ and $t_{\text{local,e}}$ are calculated by the time cost of the proposed algorithms. As a frame duration is composed of an e-slot and $T$ t-slots, the average latency \textcolor{kkb}{over} each slot in a frame duration is defined as $t_{\text{ave}}=\frac{T\times t_{\text{total,t}}+t_{\text{total,e}}}{T+1}$.} }

\textcolor{black}{Moreover, the data block of MSI is set as $B_{\text{MSI}}=n_{\text{MSI}}\times T\times B_{\text{MSI}}$~bits, where $n_{\text{MSI}}=6$ is the dimension of MSI at each slot, $T=50$ is the number of slots between the adjacent MSI exchanging, and each dimension of MSI at each slot is represented by $B_{\text{MSI}}=4$~bits. The transmission rate of lower band is set as $C_{\text{LB}}=500$~kbps~\cite{56}, the data block is set as $B_{\text{data}}=1$~Mbit, $C_{\text{ave}}$ is the average rate of mmWave band, $D_{k,\max}$ is the maximum distance between the t-UAV and the r-UAV, and $c$ is the velocity of light. As the \textcolor{kkb}{computational complexity} of \textcolor{kkb}{the} algorithms for the r-UAV is higher than that of t-UAVs, \textcolor{kkb}{the local processing time mainly depends on the time for the r-UAV to perform the beam tracking algorithms, which is estimated based on the times of multiplication and addition, and the CPU of UAVs. The CPU Intel i7-8550u~\cite{34} with processor base frequency 1.8~GHz is considered in the simulation, \textcolor{kkb}{which is adopted by a commonly-used} onboard computer ``Mainfold 2'' supporting many types of UAVs such as DJI Matrice 600 pro, DJI Matrice 600 210 series, and so on~\cite{33}.}
%and the abroad CPU. The CPU i7-8550u~[40] of the abroad computeFromr ``mainfold 2''~[39] with processor base frequency 1.8~GHz is considered in the simulation.
Note that the accurate MSI is known in e-slot by MSI exchanging and only the codeword selection without beamwidth control is performed in e-slot. The local \textcolor{kkb}{processing times in e-slot and t-slot without TE are given by} $t_{\text{local,e}}=t_{AL2,neb}+t_{AL1}$ and $t_{\text{local,t}}=t_{AL1}$, respectively, where $t_{AL2,neb}$ is the \textcolor{kkb}{computational time} of Algorithm~2 without error bounding process and $t_{AL1}$ is the \textcolor{kkb}{computational time} of Algorithm~1. When the TE is considered, $t_{\text{local,e}}=t_{AL2}+t_{AL1}$ and $t_{\text{local,t}}=t_{AL3}$, where $t_{AL2}$ is the \textcolor{kkb}{computational time} of Algorithm~2 and $t_{AL3}$ is the \textcolor{kkb}{computational time} of Algorithm~3. }

\begin{figure}[!t]
\centering
\includegraphics[scale=0.8, width=2.5in]{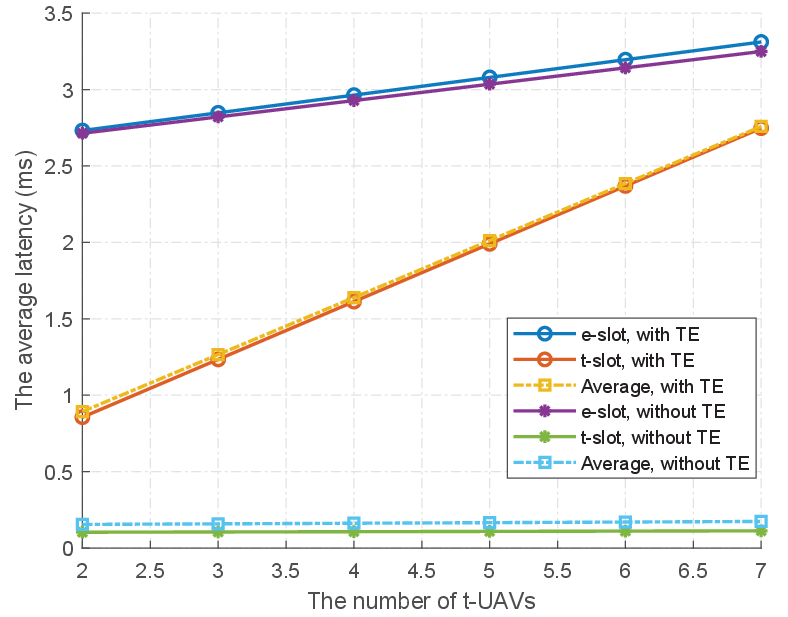}
\caption{\textcolor{black}{The average latency vs. the number of t-UAVs in e-slot, t-slot, and a frame duration consisting of an e-slot and $T$ t-slots.}}
\label{figsimulation6}
\end{figure}

\textcolor{black}{\textcolor{kkb}{The \textcolor{kkb}{schemes} with TE and without TE} are both evaluated in the simulation. In particular, the TE-aware scheme \textcolor{kkb}{is adopted for} the UAV movements with higher randomness ($\sigma^{2}=0.06$), while the scheme without TE \textcolor{kkb}{is adopted for} the UAV movements with lower randomness ($\sigma^{2}=0.05$). As shown in Fig.~\ref{figsimulation6}, the maximum average latency is less than 4~ms, \textcolor{kkb}{during which the movement distance of the UAV navigation is less than 0.08~m according to the UAV's velocity of 20~m/s~\cite{HUAC}, and hence the impact of the latency on beam angles' calculation for beam tracking can be approximately neglected. Therefore, the average latency is tolerable in the actual UAV mmWave networks. Moreover, as shown in Fig.~14, due to the relatively long interval between adjacent MSI exchanging, i.e., 500~ms ($T=50$), the average latency in a frame duration is close to that in t-slot, which is considerably lower than e-slot. Thus, the \textcolor{kkb}{proposed} algorithms can be applied to high-mobility scenarios}. }

%As the t-UAV number increases, the latency in e-slot and t-slot increases approximately linearly with the fixed subarray size which is much larger than the number of t-UAVs.

\section{Conclusion}
\label{Sec6}
In this paper, we propose a new mmWave beam tracking framework for the CA-enabled UAV mmWave network. A specialized hierarchical codebook has been constructed which fully exploits the properties of the DRE-covered CCA, wherein each codeword has a supporting subarray and the corresponding angular domain beam pattern. Then, the basic codeword selection principles have been developed for the t-UAV and the r-UAV, respectively; given the estimation of the AOA/AODs, the codeword can be quickly selected that achieves the optimized joint subarray activation and array weighting vector selection for the DRE-covered CCA. Moreover, the GP-based UAV position/attitude prediction has been proposed to track the angular information of UAV for the fast codebook-based beam tracking. The tracking error (TE) has been carefully bounded and a TE-aware codeword selection scheme has been proposed to adapt the beamwidth for better immunity against the high mobility of UAVs. Simulation results validate the performance advantages of the CA-enabled UAV mmWave network over the counterpart employing conventional UPA. Driven by our proposed codebook and codeword selection strategies, DRE-covered CCA can significantly improve the SE and coverage of the UAV mmWave network over the conventional UPA, and thus enabling agile and robust beam tracking in the highly dynamic scenarios.

% or


\begin{appendices}
\section{Proof of Theorem 1}\label{appendix:A}
The minimum number of the activated element is denoted as $n_1$ given by
%\begin{eqnarray}
%\begin{aligned}\label{P2} & \underset{\boldsymbol{f}_{{k}},{\mathcal{S}_{k}^{\text{t}}}}{\text{max}} &  & %G(\boldsymbol{f}_k\left(\mathcal{S}_{k}^{\text{t}}\right),\alpha_{t,k}(t),\beta_{t,k}(t))\\
%{\left|\boldsymbol{w}_{k}({\mathcal{S}_{k}^{\text{r}}};t)^H\boldsymbol{H}_{k}(t)\boldsymbol{f}_{k}({\mathcal{S}_{k}^{\text{t}}};t)\right|}^2\left|\boldsymbol{w}_k({\mathcal{S}_{k}^{\text{r}}})\right.\\ %\log(1+\gamma_k({\boldsymbol{f}_{{k}}({\mathcal{S}_{k}^{\text{t}}})};t)|\boldsymbol{w}_k({\mathcal{S}_{k}^{\text{r}}}))\\
% & \text{subject to} &  & \boldsymbol{f}_{{k}}\left(\mathcal{S}_{k}^{\text{t}}\right)=\boldsymbol{v}(i,j,\mathcal{S})\;\in \mathcal{V}_k.
%\end{aligned}
%\end{eqnarray}
\begin{eqnarray}
\label{n1}
\begin{aligned}
&n_{1}=\mathop{\text{arg min}}\limits_n |\alpha_0+2l\pi-\alpha_{n,\max}|\\
&\text{s.t.}\ {\alpha_0+2l\pi} \geq \alpha_{n,\max}.
\end{aligned}
\end{eqnarray}
and the maximum number of the activated element given by
\begin{eqnarray}
\label{n2}
\begin{aligned}
&n_{2}=\mathop{\text{arg min}}\limits_n |\alpha_0+2l\pi-\alpha_{n,\min}|\\
&\text{s.t.}\ {\alpha_0+2l\pi} \leq \alpha_{n,\min}.
\end{aligned}
\end{eqnarray}
Let $\alpha_0+2l\pi-\alpha_{n,\max}=0$ and substitute (\ref{DREcover}) and $\phi_c(n)$ into this equation, $n=\frac{\alpha_0+2l\pi-\Delta\alpha/2}{\Delta\phi}+\frac{N+1}{2}$. Considering the constraint in (\ref{n1}), the optimal solution $n_1$ should satisfy $\alpha_{n,\max} \leq \alpha_{n_1,\max}$. Therefore, the optimal solution of problem in (\ref{n1}) is given by $n_1=\lceil{n}\rceil$. The problem in (\ref{n2}) can be solved similarly.
\section{Proof of Theorem 2}\label{appendix:B}
%For a CCA subarray $\mathcal{S}$ with the size of $m \times n$,$m<M_{\text{act,max}}$, $n<N_{\text{act,max}}$, the sum antenna elements gain at angle $(\alpha,\beta)$ is
%The element beamwidth on the azimuth plane is given by
%\begin{eqnarray}
%BW_{a,element}=\{\alpha|\lambda_s>\rho \mathop{max}\limits_{\alpha}\ \lambda_s(\mathcal{S},\alpha,\beta_0)\},
%\end{eqnarray}
%and the element beamwidth on the azimuth plane is given by
%\begin{eqnarray}
%BW_{e,element}=\{\beta|\lambda_s>\rho \mathop{max}\limits_{\beta}\ \lambda_s(\mathcal{S},\alpha_0,\beta)\}.
%\end{eqnarray}
The element coverage for the $M \times N$-element subarray can be rewritten as
\begin{eqnarray}
\begin{aligned}
\mathcal{CV}_{\text{a(e),element}}&=\{\alpha(\beta)|\lambda_s(\alpha,\beta)>0\}\\
&=\underset{n(m)}{\cup}\{\alpha(\beta)|\left[{{\boldsymbol{\Lambda }}}\left( {{\alpha}},{\beta} \right) \right]_{(m,n)}>0\}\\
&=\underset{n(m)}{\cup}[\alpha_{n,\min}(\beta_{m,\min}),\alpha_{n,\max}(\beta_{m,\max})].\nonumber
\end{aligned}
\end{eqnarray}
When $\Delta\phi_{\text{c}}\leq\Delta\alpha$,
\begin{eqnarray}
\begin{aligned}
&\alpha_{n,\min}-\alpha_{n-1,\max}\\
&=\phi_n-\frac{\Delta\alpha}{2}+2l_1\pi-(\phi_{n-1}+\frac{\Delta\alpha}{2}+2l_2\pi)\\
&=\Delta\phi_c-\Delta \alpha+2l\pi\leq 2l\pi. \nonumber
\end{aligned}
\end{eqnarray}
Therefore,
\begin{eqnarray}
[\alpha_{n-1,\min},\alpha_{n-1,\max}]\cap[\alpha_{n,\min},\alpha_{n,\max}]=[\alpha_{n-1,\min},\alpha_{n,\max}],\nonumber
\end{eqnarray}
 and the element coverage is $\mathcal{CV}_{\text{a,element}}=[\alpha_{1,\min},\alpha_{N,\max}]$.
%\begin{eqnarray}
%\begin{aligned}
%\mathcal{CV}_{\text{a,element}}=[\alpha_{1,\min},\alpha_{N,\max}].\nonumber
%\end{aligned}
%\end{eqnarray}}

The DREs coverage of the $M \times N$-element subarray of the CCA  on the azimuth plane is
\begin{eqnarray}
\begin{aligned}
%BW_{\text{a,element}}=\alpha_{\text{max}}-\alpha_{\text{min}}+\left[\biggl\lfloor{(1-2\rho) N}\biggr\rfloor+1\right]\Delta\phi,
BW_{\text{a,element}}&=\alpha_{N,\max}-\alpha_{1,\min}\\
&=\phi_c(N)+\frac{\Delta\alpha}{2}+2l_1\pi-(\phi_c(1)-\frac{\Delta\alpha}{2}+2l_2\pi)\\
&=\Delta\alpha+\phi_c(N)-\phi_c(1)+2l\pi\\
&=\Delta\alpha+(N-1)\Delta\phi+2l\pi,\nonumber
\end{aligned}
\end{eqnarray}
As all elements on $z$-axis has the same elevation angle coverage, the DREs coverage of the subbarray is equal to the coverage of each element given by
\begin{eqnarray}
\label{elebw_ele}
BW_{\text{e,element}}=\Delta\beta.\nonumber
\end{eqnarray}
%If $\rho=\frac{1}{N}$, $BW_{\text{a,element}}=\alpha_{\text{max}}-\alpha_{\text{min}}+\left[\biggl\lfloor{2(1-\rho) N}-(N-1)\biggr\rfloor+1\right]\Delta\phi$,
\section{Proof of Theorem 3}\label{appendix:C}
According to (\ref{maxN}), when $N_{\text{act}}=N_{\text{act,max}}$, the position of the center element of all activated elements on the $xy$-plane is
%The DREs coverage of the $M \times N$-element subarray of the CCA  on the azimuth plane is
\begin{eqnarray}
\begin{aligned}
%&n_{c}^{\text{max}}=\mod\left(\biggl\lceil{\frac{\alpha}{\Delta\phi}+(N+1)/2}\biggr\rceil,N\right),\\
n_{c}^{\text{max}}&=\biggl\lceil{\frac{n_1+n_2}{2}}\biggr\rceil\\
&=\biggl\lceil\frac{1}{2}\left(\biggl\lceil{\frac{\alpha_0+2l\pi-\Delta\alpha/2}{\Delta\phi}+\frac{N+1}{2}}\biggr\rceil+\right.\\
&\left.\biggl\lceil{\frac{\alpha_0+2l\pi+\Delta\alpha/2}{\Delta\phi}+\frac{N+1}{2}}\biggr\rceil\right)\biggr\rceil\\
&=\biggl\lceil{\frac{\alpha+2l\pi}{\Delta\phi}+(N+1)/2}\biggr\rceil, l\in\mathbb{Z}.\nonumber
\end{aligned}
\end{eqnarray}
When $M_{\text{act}}=M_{\text{act,max}}$, all elements on the $z$-axis need to be activated. Therefore, the position of the center element of all activated elements on the $z$-axis is
\begin{eqnarray}
m_{c}^{\text{max}}=\biggl\lceil{\frac{1+M}{2}}\biggr\rceil.\nonumber
\end{eqnarray}
When $N_{\text{act}}<N_{\text{act,max}}$ and $M_{\text{act}}<M_{\text{act,max}}$, $n_c=n_c^{\max}$ and $m_c=m_c^{\max}$ can still allow all elements of the subarray activated as less elements need to be activated.
\textcolor{black}{
\section{Proof of Theorem 4}\label{appendix:D}
Note that as the number of the activated elements increases, the array beamwidth increases and the element beamwidth decreases. When $BW_{\text{a(e),array}}\leq BW_{\text{a(e),element}}$, $\mathcal{CV}_{\text{a(e)}}(i,j,\mathcal{S}):=\mathcal{CV}_{\text{a(e),array}}(i,j,\mathcal{S})$ and the corresponding beamwidth $BW_{\text{a(e)}}:=BW_{\text{a(e),array}}$; Otherwise, $\mathcal{CV}_{\text{a(e)}}(i,j,\mathcal{S}):=\mathcal{CV}_{\text{a(e),element}}(i,j,\mathcal{S})$ and the corresponding beamwidth $BW_{\text{a(e)}}:=BW_{\text{a(e),element}}$.
The element coverage of the $(i,j)$-the codeword in the $(m_s,n_s)$-th layer $\mathcal{CV}_{\text{a(e),element}}(i,j,\mathcal{S})$ is given by
\begin{eqnarray}
\begin{aligned}
\label{CV_element}
&\mathcal{CV}_{\text{a,element}}(i,j,\mathcal{S})\\
&=[\alpha_{1,\min}+(i-1)BW_{\text{a,element}},\alpha_{1,\min}+iBW_{\text{a,element}}],\\
&\mathcal{CV}_{\text{e,element}}(i,j,\mathcal{S})=[\beta_{\min},\beta_{\max}],
\end{aligned}
\end{eqnarray}
where $BW_{\text{a,element}}$ is given by~(\ref{elebw_az}) in Theorem~\ref{the2}.
$\mathcal{CV}_{\text{a(e),array}}(i,j,\mathcal{S})$ is the array coverage of the $(i,j)$-th codeword in the $(m_s,n_s)$-th layer, which is given by
\begin{eqnarray}
\label{CV_array}
\mathcal{CV}_{\text{a,array}}(i,j,\mathcal{S})=[(i-1)BW_{\text{a,array}},iBW_{\text{a,array}}],\\\nonumber
\mathcal{CV}_{\text{e,array}}(i,j,\mathcal{S})=[(j-1)BW_{\text{e,array}},jBW_{\text{e,array}}].
\end{eqnarray}
The array beamwidth is set as $BW_{\text{a,array}}=\frac{2\pi}{n_s}$ and $BW_{\text{e,array}}=\frac{2\pi}{m_s}$ in the $(m_s,n_s)$-th layer. When $BW_{\text{a,array}}\leq BW_{\text{a,element}}$, i.e., $\frac{2\pi}{n_s}\leq\Delta\alpha+(n_s-1)\Delta\phi$, the union of the beam coverage in the azimuth plane of all codewords in each layer is
%\begin{eqnarray}
%\begin{aligned}
%\label{CV_pro1}
%&\underset{i\in \mathcal{I}}{\cup}\mathcal{CV}_{\text{a}}(i,j,\mathcal{S})\\
%&=[0,I*BW_{\text{a,array}}]\\
%&=[0,2\pi].\\
%\end{aligned}
%\end{eqnarray}
\begin{eqnarray}
\begin{aligned}
\label{CV_pro1}
&\underset{i\in \mathcal{I}}{\cup}\mathcal{CV}_{\text{a}}(i,j,\mathcal{S})=[0,I*BW_{\text{a,array}}]=[0,2\pi].\nonumber
\end{aligned}
\end{eqnarray}
Otherwise, the union of the beam coverage is given by
\begin{eqnarray}
\begin{aligned}
\label{CV_pro2}
&\underset{i\in \mathcal{I}}{\cup}\mathcal{CV}_{\text{a}}(i,j,\mathcal{S})\\
&=[\alpha_{1,\min},\alpha_{1,\min}+I*BW_{\text{a,element}}]\\
%&\supset
&=[\alpha_{1,\min},\alpha_{1,\min}+2\pi].
\end{aligned}
\end{eqnarray}
Therefore, the union of the beam coverage of all codewords in each layer covers the whole azimuth angular domain. Revisiting Theorem~\ref{the2}, $BW_{\text{e,element}}=\Delta\beta=\pi$ and $BW_{\text{e,array}}\leq BW_{\text{e,element}}$ hold for all layers. The union of the beam coverage in the elevation plane of all codewords in each layer is
%\begin{eqnarray}
%\begin{aligned}
%\label{CV_pro3}
%&\underset{j\in \mathcal{J}}{\cup}\mathcal{CV}_{\text{e}}(i,j,\mathcal{S})\\
%&=[0,J*BW_{\text{e,array}}]\\
%&=[0,2\pi].\\
%\end{aligned}
%\end{eqnarray}
\begin{eqnarray}
\begin{aligned}
\label{CV_pro3}
&\underset{j\in \mathcal{J}}{\cup}\mathcal{CV}_{\text{e}}(i,j,\mathcal{S})=[0,J*BW_{\text{e,array}}]=[0,2\pi].\nonumber
\end{aligned}
\end{eqnarray}
Hence, the beam coverage of all codewords in each layer of the designed codebook covers the whole angular domain.
}
%&n_{c}\in\mathcal{D}_n=\left[\biggl\lceil{n_{c}^{\text{max}}-\Delta n}\biggr\rceil,\biggl\lfloor{n_{c}^{\text{max}}\Delta n}\biggr\rfloor\right],\\\label{mc}
%&m_c\in\mathcal{D}_m=\left[\biggl\lfloor{\frac{M_{\text{act}}}{2}}\biggr\rfloor,\biggl\lfloor{M-\frac{M_{\text{act}}}{2}}\biggr\rfloor\right],

%\section{Proof of Theorem 4}\label{appendix:D}
%The position of the elements in the subarray is denoted as $(m_p,n_p)$. The position of elements of the subarray is given by
%\begin{eqnarray}
%\begin{aligned}
%&(m_p,n_p)\in \mathcal{S}(m,n,m_c,n_C)\\
%&=[m_c-\frac{m}{2}, m_c+\frac{m}{2}-1]\times [n_c-\frac{n}{2}, n_c+\frac{n}{2}-1].
%\end{aligned}
%\end{eqnarray}
%If there is a conflict between the subarray $\mathcal{S}_k$ and $\mathcal{S}_q$, $\exists (m_p,n_p)$ satisfies $(m_p,n_p)\in \mathcal{S}_k$ and $(m_p,n_pp)\in \mathcal{S}_k$. Sufficiency: There are not enough number of elements to be activated by two subarraies repectively when $m_k+m_q>M$ and $\frac{n_{c,k}+n_{c,q}}{2}<d(n_{c,k}, n_{c,q})$.
%As $d(n_{c,k}, n_{c,q})$ is the least number of elements between $\mathcal{S}_k$ and $\mathcal{S}_q$ on the $xy$ plane, i
%Therefore, it is obvious that $\exists (m_{p,k},n_{p,k})\  \text{and}\  (m_{p,q},n_{p,q}), m_{p,k}=m_{p,q}\  \text{and}\  n_{p,k}=n_{p,q}$, i.e., there is a conflict between the two subarries.
\end{appendices}

%\appendices

%\section{}

%\section*{Acknowledgment}


\ifCLASSOPTIONcaptionsoff
  \newpage
\fi

\bibliographystyle{IEEEtran}%
%\bibliography{IEEEabrv,CCAmmWave3,mmWave}
\bibliography{IEEEabrv,final}

\begin{thebibliography}{10}
\providecommand{\url}[1]{#1}
\csname url@samestyle\endcsname
\providecommand{\newblock}{\relax}
\providecommand{\bibinfo}[2]{#2}
\providecommand{\BIBentrySTDinterwordspacing}{\spaceskip=0pt\relax}
\providecommand{\BIBentryALTinterwordstretchfactor}{4}
\providecommand{\BIBentryALTinterwordspacing}{\spaceskip=\fontdimen2\font plus
\BIBentryALTinterwordstretchfactor\fontdimen3\font minus
  \fontdimen4\font\relax}
\providecommand{\BIBforeignlanguage}[2]{{%
\expandafter\ifx\csname l@#1\endcsname\relax
\typeout{** WARNING: IEEEtran.bst: No hyphenation pattern has been}%
\typeout{** loaded for the language `#1'. Using the pattern for}%
\typeout{** the default language instead.}%
\else
\language=\csname l@#1\endcsname
\fi
#2}}
\providecommand{\BIBdecl}{\relax}
\BIBdecl

\bibitem{zeng2019cellular}
Y.~Zeng, J.~Lyu, and R.~Zhang, ``Cellular-connected {UAV}: Potential,
  challenges, and promising technologies,'' \emph{IEEE Wireless Commun.},
  vol.~26, no.~1, pp. 120--127, Sep. 2019.

\bibitem{zeng2019accessing}
Y.~Zeng, Q.~Wu, and R.~Zhang, ``Accessing from the sky: A tutorial on {UAV}
  communications for {5G} and beyond,'' \emph{arXiv preprint arXiv:1903.05289},
  2019.

\bibitem{55}
D.~{Ebrahimi}, S.~{Sharafeddine}, P.~{Ho}, and C.~{Assi}, ``{UAV}-aided
  projection-based compressive data gathering in wireless sensor networks,''
  \emph{{IEEE} Internet Things J.}, vol.~6, no.~2, pp. 1893 -- 1905, April
  2019.

\bibitem{30}
S.~{Yan}, M.~{Peng}, and X.~{Cao}, ``A game theory approach for joint access
  selection and resource allocation in uav assisted iot communication
  networks,'' \emph{{IEEE} Internet Things J.}, vol.~6, no.~2, pp. 1663--1674,
  2019.

\bibitem{35}
N.~{Motlagh}, T.~{Taleb}, and O.~{Arouk}, ``Low-altitude unmanned aerial
  vehicles-based internet of things services: Comprehensive survey and future
  perspectives,'' \emph{{IEEE} Internet Things J.}, vol.~3, no.~6, pp. 899 --
  922, 2016.

\bibitem{2}
T.~{Cuvelier} and R.~W. {Heath}, ``{MmWave MU-MIMO} for aerial networks,'' in
  \emph{IEEE ISWCS}, Lisbon, Portugal, Aug. 2018.

\bibitem{28}
M.~T. Dabiri, H.~Safi, S.~Parsaeefard, and W.~Saad, ``Analytical channel models
  for millimeter wave {UAV} networks under hovering fluctuations,'' \emph{arXiv
  preprint arXiv:1905.01477}, 2019.

\bibitem{32}
{X. Liu}, {Z. Li}, {N. Zhao}, {W. Meng}, {G. Gui}, {Y. Chen}, and {F. Adachi},
  ``Transceiver design and multihop {D2D} for {UAV} {IoT} coverage in
  disasters,'' \emph{{IEEE} Internet Things J.}, vol.~6, no.~2, pp. 1803--1815,
  Apr. 2019.

\bibitem{1}
Z.~{Xiao}, P.~{Xia}, and X.~{Xia}, ``Enabling {UAV} cellular with
  millimeter-wave communication: potentials and approaches,'' \emph{{IEEE}
  Commun. Mag.}, vol.~54, no.~5, pp. 66--73, May 2016.

\bibitem{zhang2019research}
C.~Zhang, W.~Zhang, W.~Wang, L.~Yang, and W.~Zhang, ``Research challenges and
  opportunities of {UAV} millimeter-wave communications,'' \emph{IEEE Wireless
  Commun.}, vol.~26, no.~1, pp. 58--62, Feb. 2019.

\bibitem{58}
``{Siklu mmWave EtherHaul-Hundred-Series products},''
  \url{https://www.sik\\lu.com/product/etherhaul-hundred-series/}.

\bibitem{13}
Z.~{Xiao}, T.~{He}, P.~{Xia}, and X.~{Xia}, ``Hierarchical codebook design for
  beamforming training in millimeter-wave communication,'' \emph{{IEEE} Trans.
  Wireless Commun.}, vol.~15, no.~5, pp. 3380--3392, May 2016.

\bibitem{14}
S.~{Park}, A.~{Alkhateeb}, and R.~W. {Heath}, ``Dynamic subarrays for hybrid
  precoding in wideband {mmWave MIMO} systems,'' \emph{{IEEE} Trans. Wireless
  Commun.}, vol.~16, no.~5, pp. 2907--2920, May 2017.

\bibitem{15}
J.~{Jin}, C.~{Xiao}, W.~{Chen}, and Y.~{Wu}, ``Channel-statistics-based hybrid
  precoding for millimeter-wave {MIMO} systems with dynamic subarrays,''
  \emph{{IEEE} Trans. Commun.}, vol.~PP, no.~99, pp. 1--1, 2019.

\bibitem{16}
C.~{Chang}, F.~{Zheng}, and S.~{Jin}, ``Fast beam training in {mmWave}
  multiuser {MIMO} systems with finite-bit phase shifters,'' in \emph{Proc.
  IEEE PIMRC}, Montreal, QC, Canada, Oct. 2017.

\bibitem{4}
L.~Josefsson and P.~Persson, \emph{Conformal array antenna theory and
  design}.\hskip 1em plus 0.5em minus 0.4em\relax John wiley \& sons, 2006,
  vol.~29.

\bibitem{8}
O.~E. {Ayach}, S.~{Rajagopal}, S.~{Abu-Surra}, Z.~{Pi}, and R.~W. {Heath},
  ``Spatially sparse precoding in millimeter wave {MIMO} systems,''
  \emph{{IEEE} Trans. Wireless Commun.}, vol.~13, no.~3, pp. 1499--1513, Mar.
  2014.

\bibitem{9}
Y.~Zeng, Q.~Wu, and R.~Zhang, ``Accessing from the sky: A tutorial on uav
  communications for 5g and beyond,'' \emph{arXiv preprint arXiv:1903.05289},
  2019.

\bibitem{6}
M.~{Xiao}, S.~{Mumtaz}, Y.~{Huang}, L.~{Dai}, Y.~{Li}, M.~{Matthaiou}, G.~K.
  {Karagiannidis}, E.~{Björnson}, K.~{Yang}, C.~{I}, and A.~{Ghosh},
  ``Millimeter wave communications for future mobile networks,'' \emph{{IEEE}
  J. Sel. Areas Commun.}, vol.~35, no.~9, pp. 1909--1935, Sep. 2017.

\bibitem{7}
Q.~{Wu}, M.~{Liu}, and Z.-R. {Feng}, ``A millimeter-wave conformal phased
  microstrip antenna array on a cylindrical surface,'' in \emph{Proc. IEEE
  APSIS}, San Diego, CA, Jul. 2008.

\bibitem{5}
M.~M. {Abdelhakam}, M.~M. {Elmesalawy}, K.~R. {Mahmoud}, and I.~I. {Ibrahim},
  ``Efficient {WMMSE} beamforming for {5G mmWave} cellular networks exploiting
  the effect of antenna array geometries,'' \emph{IET Commun.}, vol.~12, no.~2,
  pp. 169--178, Feb. 2018.

\bibitem{26}
S.~{He}, J.~{Wang}, Y.~{Huang}, B.~{Ottersten}, and W.~{Hong}, ``Codebook-based
  hybrid precoding for millimeter wave multiuser systems,'' \emph{{IEEE} Trans.
  Signal Process.}, vol.~65, no.~20, pp. 5289--5304, Oct 2017.

\bibitem{27}
J.~{Singh} and S.~{Ramakrishna}, ``On the feasibility of codebook-based
  beamforming in millimeter wave systems with multiple antenna arrays,''
  \emph{{IEEE} Trans. Wireless Commun.}, vol.~14, no.~5, pp. 2670--2683, May
  2015.

\bibitem{11}
K.~{Chen} and C.~{Qi}, ``Beam training based on dynamic hierarchical codebook
  for millimeter wave massive {MIMO},'' \emph{{IEEE} Commun. Lett.}, vol.~23,
  no.~1, pp. 132--135, Jan. 2019.

\bibitem{12}
R.~{Zhang}, H.~{Zhang}, W.~{Xu}, and C.~{Zhao}, ``A codebook based simultaneous
  beam training for mmwave multi-user {MIMO} systems with split structures,''
  in \emph{Proc. IEEE GLOBECOM}, Abu Dhabi, United Arab Emirates, Dec. 2018.

\bibitem{23}
J.~{Zhao}, F.~{Gao}, W.~{Jia}, S.~{Zhang}, S.~{Jin}, and H.~{Lin}, ``Angle
  space channel tracking for hybrid {mmWave} massive {MIMO} systems,'' in
  \emph{Proc. IEEE GLOBECOM}, Singapore, Dec. 2017.

\bibitem{20}
I.~{Mavromatis}, A.~{Tassi}, R.~J. {Piechocki}, and A.~{Nix}, ``Beam alignment
  for millimetre wave links with motion prediction of autonomous vehicles,'' in
  \emph{Antennas, Propagation RF Technology for Transport and Autonomous
  Platforms}, Birmingham, UK, Feb. 2017.

\bibitem{21}
L.~{Dai} and X.~{Gao}, ``Priori-aided channel tracking for millimeter-wave
  beamspace massive {MIMO} systems,'' in \emph{Proc. IEEE URSI AP-RASC}, Seoul,
  South Korea, Aug. 2016.

\bibitem{19}
J.~{Zhao}, F.~{Gao}, Q.~{Wu}, S.~{Jin}, Y.~{Wu}, and W.~{Jia}, ``Beam tracking
  for {UAV} mounted satcom on-the-move with massive antenna array,''
  \emph{{IEEE} J. Sel. Areas Commun.}, vol.~36, no.~2, pp. 363--375, Feb. 2018.

\bibitem{22}
J.~{Zhao}, F.~{Gao}, L.~{Kuang}, Q.~{Wu}, and W.~{Jia}, ``Channel tracking with
  flight control system for {UAV mmWave MIMO} communications,'' \emph{{IEEE}
  Commun. Lett.}, vol.~22, no.~6, pp. 1224--1227, Jun. 2018.

\bibitem{25}
{J. {Zhang}}, {W. {Xu}}, {H. {Gao}}, {M. {Pan}}, {Z. {Feng}}, and {Z. {Han}},
  ``Position-attitude prediction based beam tracking for {UAV} mmwave
  communication,'' in \emph{Proc. IEEE ICC}, Shanghai, China, May 2019.

\bibitem{xu2008pattern}
Z.~Xu, H.~Li, Q.~Liu, and J.~Li, ``Pattern synthesis of conformal antenna array
  by the hybrid genetic algorithm,'' \emph{Progress In Electromagnetics
  Research}, vol.~79, pp. 75--90, 2008.

\bibitem{SpatiallySparse}
O.~E. Ayach, S.~Rajagopal, S.~Abu-Surra, Z.~Pi, and R.~W. Heath, ``Spatially
  sparse precoding in millimeter wave {MIMO} systems,'' \emph{{IEEE} Trans.
  Wireless Commun.}, vol.~13, no.~3, pp. 1499--1513, Mar. 2014.

\bibitem{Smooth-Turn}
Y.~Wan, K.~Namuduri, Y.~Zhou, and S.~Fu, ``A {Smooth-Turn} mobility model for
  airborne networks,'' \emph{{IEEE} Trans. Veh. Technol.}, vol.~62, no.~7, pp.
  3359--3370, Sep. 2013.

\bibitem{attitude}
D.~Kingston and R.~Beard, ``Real-time attitude and position estimation for
  small {UAVs} using low-cost sensors,'' in \emph{Proc. AIAA 3rd "Unmanned
  Unlimited" Technical Conference, Workshop and Exhibit}, Chicago, IL, Apr.
  2004.

\bibitem{lowfre}
D.~Ramasamy, S.~Venkateswaran, and U.~Madhow, ``Compressive tracking with
  1000-element arrays: A framework for multi-gbps mm wave cellular downlinks,''
  in \emph{Proc. 50th Annual Allerton Conference on Communication, Control, and
  Computing (Allerton)}, Monticello, IL, Oct. 2012.

\bibitem{HUAC}
``\textcolor{black}{The details of {CHCNAV UAV platform}},''
  \url{\textcolor{black}{https://www.chcnav.com/\\products/mobile-mapping-systems/uav-platforms}}.

\bibitem{56}
{T. Andre}, {K. Hummel}, {A. Schoellig}, {E. Yanmaz}, {M. Asadpour}, {C.
  Bettstetter}, {P. Grippa}, {H. Hellwagner}, {S. Sand}, and {S. Zhang},
  ``Applicationdriven design of aerial communication networks,'' \emph{{IEEE}
  Commun. Mag.}, vol.~52, no.~5, pp. 129--137, May 2014.

\bibitem{34}
``{Intel@Core} i7-8550u processor,''
  \url{https://www.intel.com/content/\\www/us/en/products/processors/core/i7-processors/i7-8550u.html}.

\bibitem{33}
``{DJI Manifold} 2,'' \url{https://www.dji.com/manifold-2/specs}.

\end{thebibliography}
%\vspace*{-3\baselineskip}
\begin{IEEEbiography}[{\includegraphics[width=1in,height=1.25in,clip,keepaspectratio]{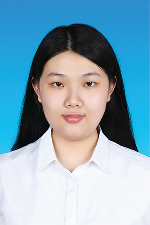}}]{Jinglin Zhang}
received her B.S. degree in communication engineering from Beijing University of Posts and Telecommunications (BUPT), China, in 2017. She is currently pursuing the Ph.D. degree in information and communication engineering from BUPT, China.
Her research interests include mmWave communications and networks, UAV communications and networks, and machine learning. She was a recipient of the Best Paper Award in IEEE ICC 2019.
\end{IEEEbiography}
%\vspace*{-5\baselineskip}
\begin{IEEEbiography}[{\includegraphics[width=1in,height=1.25in,clip,keepaspectratio]{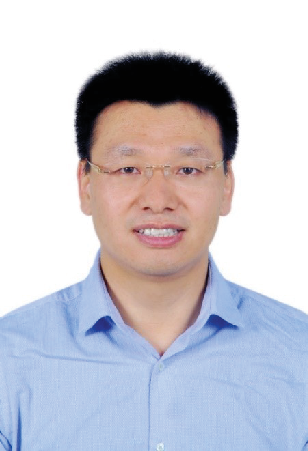}}]{Wenjun Xu}
is a professor and Ph.D. supervisor in School of Information and Communication Engineering at Beijing University of Posts and Telecommunications (BUPT), Beijing, China. He received his B.S. and Ph.D. degrees from BUPT, in 2003 and 2008, respectively. He currently serves as a center director of the Key Laboratory of Universal Wireless Communications, Ministry of Education, P. R. China. He is a senior member of IEEE, and is now an Editor for China Communications. His research interests include AI-driven networks, UAV communications and networks, green communications and networking, and cognitive radio networks.
\end{IEEEbiography}
%\vspace*{-4\baselineskip}
\begin{IEEEbiography}[{\includegraphics[width=1in,height=1.25in,clip,keepaspectratio]{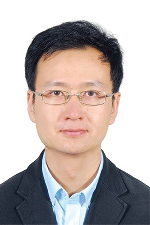}}]{Hui Gao}
 (S'10-M'13-SM'16) received the B.Eng. degree in information engineering and the Ph.D. degree in signal and information processing from the Beijing University of Posts and Telecommunications (BUPT), Beijing, China, in 2007 and 2012, respectively. From 2009 to 2012, he was a Research Assistant with the Wireless and Mobile Communications Technology Research and Development Center, Tsinghua University, Beijing. In 2012, he was a Research Assistant with the Singapore University of Technology and Design, Singapore, where he was a Postdoctoral Researcher, from 2012 to 2014. He is currently an Associate Professor with the School of Information and Communication Engineering, BUPT. His research interests include massive and mmWave MIMO systems, UAV communications, intelligent reflecting surface aided wireless network.
\end{IEEEbiography}
\begin{IEEEbiography}[{\includegraphics[width=1in,height=1.25in,clip,keepaspectratio]{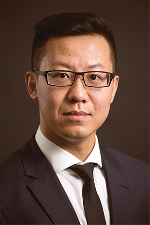}}]{Miao Pan}
 (S'07-M'12-SM'18) received his BSc degree in Electrical Engineering from Dalian University of Technology, China, in 2004, MASc degree in electrical and computer engineering from Beijing University of Posts and Telecommunications, China, in 2007 and Ph.D. degree in Electrical and Computer Engineering from the University of Florida in 2012, respectively. He is now an Associate Professor in the Department of Electrical and Computer Engineering at University of Houston. He was a recipient of NSF CAREER Award in 2014. His research interests include cybersecurity, big data privacy, deep learning privacy, cyber-physical systems, and cognitive radio networks. His work won IEEE TCGCC (Technical Committee on Green Communications and Computing) Best Conference Paper Awards 2019, and Best Paper Awards in ICC 2019, VTC 2018, Globecom 2017 and Globecom 2015, respectively. Dr. Pan is an Editor for IEEE Open Journal of Vehicular Technology and an Associate Editor for IEEE Internet of Things (IoT) Journal (Area 5: Artificial Intelligence for IoT), and used to be an Associate Editor for IEEE Internet of Things (IoT) Journal (Area 4: Services, Applications, and Other Topics for IoT) from 2015 to 2018. He has also been serving as a Technical Organizing Committee for several conferences such as TPC Co-Chair for Mobiquitous 2019, ACM WUWNet 2019. He is a member of AAAI, a member of ACM, and a senior member of IEEE.
\end{IEEEbiography}
%\vspace*{-3\baselineskip}
\begin{IEEEbiography}[{\includegraphics[width=1in,height=1.25in,clip,keepaspectratio]{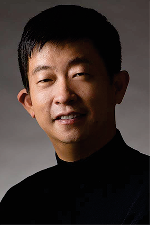}}]{Zhu Han}
  (S'01-M'04-SM'09-F'14) received the B.S. degree in electronic engineering from Tsinghua University, in 1997, and the M.S. and Ph.D. degrees in electrical and computer engineering from the University of Maryland, College Park, in 1999 and 2003, respectively.

From 2000 to 2002, he was an R\&D Engineer of JDSU, Germantown, Maryland. From 2003 to 2006, he was a Research Associate at the University of Maryland. From 2006 to 2008, he was an assistant professor at Boise State University, Idaho. Currently, he is a John and Rebecca Moores Professor in the Electrical and Computer Engineering Department as well as in the Computer Science Department at the University of Houston, Texas. His research interests include wireless resource allocation and management, wireless communications and networking, game theory, big data analysis, security, and smart grid.  Dr. Han received an NSF Career Award in 2010, the Fred W. Ellersick Prize of the IEEE Communication Society in 2011, the EURASIP Best Paper Award for the Journal on Advances in Signal Processing in 2015, IEEE Leonard G. Abraham Prize in the field of Communications Systems (best paper award in IEEE JSAC) in 2016, and several best paper awards in IEEE conferences. Dr. Han was an IEEE Communications Society Distinguished Lecturer from 2015-2018, AAAS fellow since 2019 and ACM distinguished Member since 2019. Dr. Han is 1\% highly cited researcher since 2017 according to Web of Science.
\end{IEEEbiography}
%\vspace*{-3\baselineskip}
\begin{IEEEbiography}[{\includegraphics[width=1in,height=1.25in,clip,keepaspectratio]{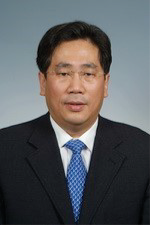}}]{Ping Zhang}
 (M'07-SM'15-F'18) is currently a professor of School of Information and Communication Engineering at Beijing University of Posts and Telecommunications, the director of State Key Laboratory of Networking and Switching Technology, a member of IMT-2020 (5G) Experts Panel, a member of Experts Panel for China 6G development. He served as Chief Scientist of National Basic Research Program (973 Program), an expert in Information Technology Division of National High-tech R\&D program (863 Program), and a member of Consultant Committee on International Cooperation of National Natural Science Foundation of China. His research interests mainly focus on wireless communication. He is an Academician of the Chinese Academy of Engineering (CAE).
\end{IEEEbiography}
\end{document}